\documentclass[11pt]{article}
\usepackage{jheppub}
\usepackage{mathtools}
\usepackage{amssymb}
\usepackage{amsfonts}
\usepackage{latexsym}
\usepackage{cancel}
\usepackage{pdfpages}
\usepackage{graphicx,color}

\def\cA{{\cal A}}

\def\cD{{\cal D}}
\def\cC{{\cal C}}

\def\cF{{\cal F}}

\def\cH{{\cal H}}

\def\cK{{\cal K}}
\def\cL{{\cal L}}

\def\cN{{\cal N}}
\def\cO{{\cal O}}
\def\cP{{\cal P}}
\def\cR{{\cal R}}
\def\cT{{\cal T}}
\def\cU{{\cal U}}

\def\cW{{\cal W}}

\newbox\charbox
\newbox\slabox
\def\slsh#1{{      
        \setbox\charbox=\hbox{$#1$}
        \setbox\slabox=\hbox{$/$}
        \dimen\charbox=\ht\slabox
        \advance\dimen\charbox by -\dp\slabox
        \advance\dimen\charbox by -\ht\charbox
        \advance\dimen\charbox by \dp\charbox
        \divide\dimen\charbox by 2
        \raise-\dimen\charbox\hbox to \wd\charbox{\hss/\hss}
        \llap{$#1$}
}}

\def\exd{{\hbox{d}}}

\def\cA{{\cal A}}

\def\cD{{\cal D}}
\def\cC{{\cal C}}

\def\cF{{\cal F}}
\def\cH{{\cal H}}

\def\cL{{\cal L}}

\def\cO{{\cal O}}

\def\cR{{\cal R}}
\def\cS{{\cal S}}
\def\cT{{\cal T}}
\def\cU{{\cal U}}
\def\cW{{\cal W}}
\def\cP{{\cal P}}

\def\Tr{{\rm Tr}}

\def\bea{\begin{eqnarray}}
\def\eea{\end{eqnarray}}
\def\be{\begin{equation}}
\def\ee{\end{equation}}

\def\ssA{{\scriptscriptstyle A}}
\def\ssB{{\scriptscriptstyle B}}

\def\IR{{\scriptscriptstyle IR}}

\def\AB{{\ssA\ssB}}

\def\nn{\nonumber}

\def\({\left(}
\def\){\right)}

\def\pref#1{(\ref{#1})}

\title{Open EFTs, IR Effects \& Late-Time Resummations:\\
Systematic Corrections in Stochastic Inflation }

\author[a,b]{C.P.~Burgess,\,}
\affiliation[a]{Physics \& Astronomy, McMaster University,
Hamilton, ON, Canada, L8S 4M1}
\affiliation[b]{Perimeter Institute for Theoretical Physics, Waterloo, ON, Canada N2L 2Y5}
\author[c]{ R.~Holman, }
\affiliation[c]{ Physics Department, Carnegie Mellon University, Pittsburgh PA
15213 USA}
\author[d]{ and G.~Tasinato}
\affiliation[d]{Department of Physics, Swansea University, Swansea, SA2 8PP, U.K.}
\abstract{Though simple inflationary models describe the CMB well, their corrections are often plagued by infrared effects that obstruct a reliable calculation of late-time behaviour. We adapt to cosmology tools designed to address similar issues in other physical systems with the goal of making reliable late-time inflationary predictions. The main such tool is Open EFTs which reduce in the inflationary case to Stochastic Inflation plus calculable corrections. We apply this to a simple inflationary model that is complicated enough to have dangerous IR behaviour yet simple enough to allow the inference of late-time behaviour. We find corrections to standard Stochastic Inflationary predictions for the noise and drift, and we find these corrections ensure the IR finiteness of both these quantities. The late-time probability distribution, $\cP(\phi)$, for super-Hubble field fluctuations are obtained as functions of the noise and drift and so these too are IR finite. We compare our results to other methods (such as large-$N$ models) and find they agree when these models are reliable. In all cases we can explore in detail we find IR secular effects describe the slow accumulation of small perturbations to give a big effect: a significant distortion of the late-time probability distribution for the field. But the energy density associated with this is only of order $H^4$ at late times and so does {\em not} generate a dramatic gravitational back-reaction.
}

\begin{document}
\maketitle

\flushbottom

\section{Introduction}

Precision CMB measurements reveal a remarkable pattern of primordial correlations over large scales. Part of the appeal of inflationary models is their ability to explain these as vacuum fluctuations enormously stretched by universal expansion until writ large across the sky \cite{CMBth1}. The vacuum fluctuations used for this purpose are essentially those of free massless fields in de Sitter space, as are believed to dominate in the weak-field regime of central interest for most slow-roll models. Implicit in this belief is that any weak interactions present can be neglected to leading order in a controlled approximation. 

This picture is undermined by explicit calculations of perturbations within near-de Sitter geometries. As has long been known \cite{InfPred,  StochInf}, these generically reveal two related problems \cite{IRdiv}. The first is the infrared (IR) singularity of many quantities of interest (such as $n$-point field correlations) and the second is the presence of `secular' evolution (see {\em e.g.}~the review \cite{seery-rev}), for which powers of perturbative couplings arise systematically multiplied by powers of $\ln a = Ht$. The first problem signals the importance of long-wavelength modes to making predictions and the second causes perturbation theory to fail at late times. Although in single field models such IR problems are plausibly gauge artefacts \cite{gaugeart} (see also the review \cite{Tanaka:2013caa}), this need not be true in more general models so their presence threatens the perturbative control required to exclude large theoretical uncertainties in predicting observable implications for the post-inflationary universe \cite{IRdiv, PtbnFailure}. Some argue this breakdown suggests the development of a large back-reaction that might indicate an instability of de Sitter space itself \cite{IRdiv, dSunstable}.

IR issues are most transparent when expressed within the effective field theory (EFT) of the longest-wavelength modes. We here follow \cite{OpenEFT} (see also \cite{OpenEFTother}) and identify the relevant long-wavelength EFT for cosmology using the language of open systems, a research area started with \cite{Feynman:1963fq} (see also \cite{Calzetta:1999zr} for a review on applications to cosmology). Because super-Hubble modes move through an environment of sub-Hubble modes with which information is exchanged (such as when modes pass from sub- to super-Hubble at horizon exit) they form an open system. Consequently their effective description is less like EFTs encountered elsewhere in gravity \cite{EFTs} and cosmology \cite{TPI, InfEFT} than it is like the effective description of a particle moving through a fluid.\footnote{The open nature of the problem shares some features of -- but is not equivalent to -- an effective description of the cosmic fluid, such as that described in \cite{LSSEFT}.} And like for a particle interacting with an environment it is generic that even very small interactions can accumulate to cause large effects at late times since the environment never goes away; no matter how small the interaction, $V$, the evolution operator $e^{-iVt}$ is eventually not close to unity. 

Experience with similar problems in non-gravitational settings suggests the key tool for resumming late-time predictions starting from perturbative interactions is the coarse-grained master equation that describes the evolution of the density matrix for the long-wavelength part of the system that is of interest \cite{OpenSys}. This master equation is obtained from the Liouville equation by tracing out irrelevant short-distance modes and (as described in \cite{OpenEFT}) when applied to inflationary cosmology the leading contribution for super-Hubble modes gives Starobinsky's stochastic inflation \cite{StochInf}. Subleading interactions describe various corrections including a description of the decoherence of the super-Hubble modes by their shorter-wavelength brethren. 

Because stochastic inflation arises as the leading approximation to a broader formalism designed to resum late-time effects, one might expect stochastic calculations to resolve some or all of the IR issues in cosmology. There is indeed evidence that this is true in several simple examples \cite{TsamisWoodard}, such as for a spectator scalar field in de Sitter space subject to a $\lambda \phi^4$ interaction. We here build yet more evidence for this using a toy system that is complicated enough to display IR and secular effects, but simple enough to solve explicitly to extract reliably late-time evolution. 

We start, in \S\ref{section:lindblad}, by reviewing briefly how master-equation techniques can be used to extend perturbative calculations reliably to very late times. (Such arguments underlie, for example, the ability to compute an index of refraction relevant to the geometrical optics limit, despite the breakdown of naive perturbative techniques for the photon-atom interactions well before this time.) This section also briefly recaps the stochastic limit in cosmology and summarizes evidence for their relevance to late-time evolution in $\lambda \phi^4$ theory.

The toy model of interest is defined in \S\ref{section:future}. We work with a multi-field inflationary picture in order to circumvent no-go arguments specific to single-field models. For simplicity we specialize to the case of a spectator scalar (or scalars) whose energy density plays no role in driving the universal expansion. For such a scalar we investigate a three-parameter deformation from a massless spectator scalar in de Sitter space:\footnote{Appendix \ref{app:ghostinf} extends the discussion to include non-standard dispersion relations.} a free spectator scalar field (or, sometimes, $N$ scalars) with mass, $m$, time-dependent speed of sound, $c_s$, within power-law inflation (with constant slow-roll parameter $\epsilon = - \dot H/H^2$). On one hand the model is exactly solvable and its late-time behaviour can be exactly obtained; on the other hand it exhibits IR singularities and secular issues when the parameters $m/H$, $s=\exd c_s/\exd \ln t$ and $\epsilon$ are perturbatively small.  

Comparing the perturbative and exact solutions yields the following results:
\begin{itemize}
\item We construct the system's mode functions and use these to compute explicitly how the mean, $\langle \phi \rangle$, and variance, $\langle \phi^2 \rangle - (\langle \phi \rangle)^2$, of the super-Hubble modes of the field evolve. We then use these to identify the equivalent Fokker-Planck equation describing the evolution of the corresponding probability distribution, $P(\phi,t)$ (and while doing so simplify the arguments given for its derivation in \cite{OpenEFT}). 
\item In the naive derivation we compute the noise and drift coefficients, $\cN$ and $\cF$, as a function of the three parameters ($m/H$, $s$ and $\epsilon$) as well as time. Very little must be assumed about the time-evolution of the state in this calculation, but the noise, $\cN$, in general also inherits the IR divergence and secular effects that are found in $\partial_t \langle \phi^2 \rangle$.
\item A better derivation of the Fokker-Planck equation instead identifies $\cN$ and $\cF$ as functions of $m/H$, $s$ and $\epsilon$ and the field $\phi$ rather than time. This is better because it is in this form that the Fokker-Planck equation can be integrated to obtain the late-time limit $\cP(\phi) = \lim_{t\to \infty} P(\phi,t)$ as a function of $\cN$ and $\cF$. We perform this calculation and find it reproduces standard expressions --- $\cN = H^3/8\pi^2$ and $\cF = V'/3H$ --- to leading order in $m/H$, $s$ and $\epsilon$. Most importantly, however, we also find subdominant corrections to both $\cN$ and $\cF$ as functions of $m/H$, $s$ and $\epsilon$. 
\item In general we find that although correlation functions like $\langle \phi^{2} \rangle$ diverge in the IR for some choices of $m$ and $\epsilon$, because of the corrections mentioned in the previous bullet point these divergences precisely cancel to give an IR finite noise and drift, $\cN$ and $\cF$. The IR finiteness of $\cN$ and $\cF$ is consistent with IR singularities in $\langle \phi^2 \rangle$ because these singularities arise from singularities in the {\em fluctuations} of $\cF$: $\langle \phi \cF \rangle - \langle \phi \rangle \langle \cF \rangle$.
\item The IR finiteness of $\cN$ and $\cF$ ensures that the late-time solution, $\cP(\phi)$, of the Fokker-Planck equation is also IR finite. This is useful since it is likely a prerequisite for proving more generally the IR finiteness of late-time observables. 
\item Our results for free massive fields can be used in whole cloth to compute the noise, drift and late-time distributions for $N$ scalars interacting through a quartic $\lambda \phi^4$ interaction in the large-$N$ limit, and we compute these as a function of $g = \lambda N$ for this system. We show how the late time result agrees in this case with results of other methods in the large-$N$ limit, in cases where these are known. 
\item We generalize our derivation to include the case where the scalar mass depends on its vev and so also slowly changes in time. By doing so we derive the late-time limit of spectator scalars self-interacting through a quartic $\lambda \phi^4$ interaction, without making recourse to the large-$N$ limit. We find a result that approaches the standard Starobinsky result plus corrections, that disagrees with what would be obtained for this system using the Hartree approximation.
\end{itemize} 

In the cases where we can compute the late-time limit we find secular evolution does indeed accumulate to cause relatively large effects at late times. Usually the large effect is a significant distortion of the late-time probability away from the initially gaussian distribution experienced by each mode as it crosses the horizon. In no cases did we find an equally large accumulation of energy density and gravitational back-reaction, with the super-Hubble contribution to the stress-energy remaining only of order $H^4$ at late times. Consequently in none of our examples does secular evolution indicate an incipient instability of de Sitter space. 

We note that Langevin type equations can also appear if one considers a rolling inflaton coupled to other scalars and then integrates these scalars out. This was done in refs.\cite{BoyanEFT}

Our conclusions are briefly summarized in \S\ref{section:discussion}, with a short outline of possible future directions. Various appendixes contain technical details and extensions of the arguments used in the main text. 

\section{IR singularities, secular evolution and resummation}
\label{section:lindblad}

This section is meant to summarize two results. We first lay out the general case as to why stochastic arguments should be expected to resum secular evolution and so to capture the late-time evolution of inflationary perturbations. Following \cite{OpenEFT} this is done by showing it to be a special case of a more general technique widely used outside of cosmology to resum secular effects. We then briefly summarize the present concrete evidence for this argument, coming from the explicit inflationary calculations of \cite{TsamisWoodard}.

\subsection{Stochastic inflation: the cosmic master (equation)}
\label{ssec:stochasticlatetime}

Why should stochastic methods be related to IR singularities and secular evolution in cosmology? The starting point is the recognition that the basic problem is the breakdown of perturbation theory at late times, and that this problem also arises (and has been solved) in many other areas of physics. Master-equation methods are among the tools developed to deal with this problem, and we here repeat the case made in \cite{OpenEFT} that these methods reduce to those of Stochastic Inflation (plus systematic corrections) when applied in an inflationary context. To do so we first give a very brief recap of Master-equation methods in general \cite{OpenSys}, followed by a statement of their implications for simple inflationary models. 

\subsubsection*{Interacting open systems}
\label{sssec:opensys}

The basic problem arises whenever two systems --- call them $A$ and $B$ --- interact over arbitrarily long times. Given a hamiltonian of the form $H = H_\ssA + H_\ssB + H_{\ssA\ssB}$ consisting of terms that evolve $A$ and $B$ separately plus an interaction between them, no matter how small the interaction $H_{\ssA\ssB}$ is there is always a time, $t_p$, beyond which it is a bad approximation to evaluate $\exp[- i H_{\ssA\ssB} t]$ in powers of $H_\AB$. In this sense it is generic that perturbative late-time predictions can be problematic whenever interactions do not turn off with time.\footnote{The scattering problems studied in introductory courses on quantum field theory are among the few cases where this is {\em not} an issue because the separation of particle wavepackets turns off the mutual interactions.}

Our interest is situations where all measurements are performed exclusively on system $A$ and predictions are sought on how their late-time results are influenced by the presence of sector $B$. It is useful to have a concrete example in mind when describing the formalism, such as the interactions of a particle (sector $A$) traveling through a medium (sector $B$) --- perhaps a photon within a transparent material or a neutrino passing through the Sun. In general knowing the evolution of any observable, $\cA(t) = \Tr(\cO_\ssA)$, involving only sector $A$ is equivalent to knowing the evolution of the reduced density matrix, $\rho_\ssA(t) = \Tr_\ssB[ \rho(t) ]$, obtained by tracing the full density matrix, $\rho$, over the unobserved sector $B$, because for such observables $\cA(t) = \Tr_\ssA[ \rho_\ssA(t) \cO_\ssA ]$.

In principle, the evolution $\rho_\ssA(t)$ is completely governed by the evolution of $\rho(t)$, which in the interaction picture is obtained by solving the Liouville equation
\be \label{LiouvilleFull}
 \frac{\partial \rho}{\partial t} = -i \, \Bigl[ H_\AB \,, \rho  \Bigr]  \,,
\ee
and so has the familiar perturbative solution
\bea \label{LiouvillePSoln}
 \rho(t) &=& \rho_0 - i \int_0^t \exd \tau \; \Bigl[ H_\AB(\tau) \,, \rho_0   \Bigr] + (-i)^2 \int_0^t \exd \tau \int_0^\tau \exd \tilde\tau \Bigl[ H_\AB(\tilde \tau) \,,  \Bigl[ H_\AB(\tau) \,, \rho_0  \Bigr] \Bigr] + \cdots \nn\\
 &=& \left\{ \cT \; \exp\left[ - i  \int_0^t \exd \tau \; H_\AB(\tau) \right] \right\}  \rho_0 \left\{ \cT \; \exp \left[ -i  \int_0^t \exd \tilde \tau \; H_\AB(\tilde \tau) \right] \right\}^* \,,
\eea
where $H_\AB(t) := \exp(i H_0 t) H_\AB \exp(-i H_0 t)$ with $H_0 := H_\ssA + H_\ssB$, $\cT$ denotes the appropriate time-ordering of the integrals and so on. This explicitly shows the potential problem with perturbative methods if the integrands do not vanish quickly enough at large times.

In general solving the equation that results for $\rho_\ssA$ is a mess, particularly at late times. However relative simplicity can occur if: ($i$) the system starts in an initially uncorrelated state, $\rho_0 = \varrho_\ssA \otimes \varrho_\ssB$; and ($ii$) the autocorrelation function of $H_\AB$ in sector $B$ vanishes for large enough times --- that is if there exists a $t_c$ for which
\be
 \Bigl \langle \delta H_\AB(t) \, \delta H_\AB(t') \Bigr \rangle_\ssB \to 0 \qquad \hbox{for $t \gg t_c$} \,,
\ee
where $\delta H_\AB := H_\AB - \langle H_\AB \rangle_\ssB$ and $\langle \cdots \rangle_\ssB := \Tr_\ssB[ \varrho_\ssB \, H_\AB ]$. The simplicity arises because the correlations between systems $A$ and $B$ become less and less important for the evolution of $A$ over times much longer than $t_c$, allowing an approximate description that effectively expands in the autocorrelations of $H_\AB$. 

This mean-field/fluctuation split is most efficiently implemented in terms of the full evolution operator, $U(t) = \cT \exp \left[ -i  \int_0^t \exd \tau \; H_\AB (\tau) \right]$, as follows:
\be
 U(t) =: \overline U(t) + \cU(t) \,,
\ee
where $\overline U(t) := \langle U(t) \rangle_\ssB = \Tr_\ssB [ \varrho_\ssB \, U(t) ]$, because then the condition $\langle \,\cU(t) \rangle_\ssB = 0$ ensures the evolution of $\rho_\ssA(t)$ nicely splits into a `mean' and `fluctuation' part, with no cross terms:
\be
  \rho_\ssA(t) = \Tr_\ssB[ U(t) \, \rho_0 \, U^*(t) ] =  \overline U(t) \, \varrho_\ssA \, \overline U^*(t) + \Tr_\ssB [ \,\cU(t) \, \rho_0 \, \cU^*(t) ] \,.
\ee
The mean Hamiltonian is then defined by $\overline U =: \cT  \exp \left[ -i \int_0^t \exd \tau \; \overline H (\tau) \right]$, or equivalently
\be \label{MeanHdef}
 \overline H = i \left( \frac{\partial \overline U}{\partial t} \right) \; \overline U^{-1}  
 = \langle H_\AB \rangle_\ssB - i \int_0^t \exd \tau \;   \Bigl\langle \delta H_\AB(t) \, \delta H_\AB(\tau) \Bigr\rangle_\ssB + \cdots \,, 
\ee
and so on. 

For the concrete case of light interacting with a polarizable medium it is $\overline U$ that describes the coherent evolution (with the second term in \pref{MeanHdef} turning out to be responsible for the index of refraction), while $\cU$ describes the incoherent `diffuse' scattering that can make a medium opaque. (Since both arise at second order in $H_\AB$ a large-$N$ argument is required to allow materials to be transparent while still having an index of refraction not too close to 1.) Similarly it is $\langle H_\AB \rangle_\ssB$ that describes the medium-dependent interactions responsible for MSW oscillations within the Sun \cite{MSW}, while the terms quadratic in $H_\AB$ give the leading deviations \cite{MSWdev, MSWexp} from the MSW approximation. (For neutrinos there is no particular utility in distinguishing $\overline U$ from $\cU$ at second order because of the comparatively short neutrino wavelength and the very feeble nature of the interactions.)

\subsubsection*{Master-equation methods}
\label{sssec:mastereqmethods}

Nothing said so far directly addresses the issue of making late-time predictions using perturbative methods. Progress on this is possible if there is a hierarchy between the characteristic times: $t_c \ll t_p$, because when this is true it is possible to define a `coarse-grained' evolution for $\rho_\ssA(t)$:
\bea \label{coarse}
 \frac{D \rho_\ssA}{Dt} &:=& \frac{1}{\Delta t} \Bigl[ \rho_\ssA(t+\Delta t) - \rho_\ssA(t) \Bigr] \nn\\
 &=& \frac{1}{\Delta t} \Tr_\ssB \Bigl[ \overline U(\Delta t) \, \rho_\ssA(t) \, \overline U^*(\Delta t) \Bigr] + \frac{1}{\Delta t} \Tr_\ssB \Bigl[ \cU(\Delta t) \, \rho(t) \, \cU^*(\Delta t) \Bigr] \\
 &=&  - \frac{i}{\Delta t} \int_t^{t+\Delta t} \exd \tau \Bigl[ \langle H_\AB(\tau) \rangle_\ssB \,, \rho_\ssA (t)\Bigr]  + \cdots  \,, \nn  
\eea
where the ellipses represent terms at least second-order in $H_\AB$.

The assumed hierarchy allows the choice $t_c \ll \Delta t \ll t_p$, which we now make. On one hand the inequality $\Delta t \ll t_p$ ensures the integration over $\tau$ does not ruin the validity of perturbing in $H_\AB$. On the other hand the inequality $t_c \ll \Delta t$ means that the right-hand side of \pref{coarse} can `forget' the correlations between $A$ and $B$, potentially allowing a dependence on the instantaneous value of $\rho$ rather than on the entire history of what happened within the interval $(t, t+ \Delta t)$. If so \pref{coarse} can be written schematically as
\be \label{coarseschem}
 \frac{D \rho_\ssA}{Dt} = \cF(\rho_\ssA, \rho_\ssB) = \sum_{k=1}^\infty \cF_k(\rho_\ssA, \rho_\ssB) \,, 
\ee
where $\cF$ is a calculable function that may be evaluated perturbatively in $H_\AB$ (with $\cF_k$ denoting the contribution at $k$-th order). Given a specific function $\cF$ one can read \pref{coarseschem} as a differential equation to be solved for $\rho_\ssA$ (and possibly also $\rho_\ssB$ if sector $B$ also evolves in response to $A$). 

Now comes the main point. The requirement $\Delta t \ll t_p$ might lead one to think that no progress has been made on learning the late-time behaviour, but this is incorrect. Solutions to \pref{coarseschem} can be trusted even for times $t \gg t_p$, provided \pref{coarseschem} itself is valid for a window of width $\Delta t$ around {\em any} specific $t$. Solutions found by integrating remain valid so long as an overlapping set of windows of width $\Delta t$ exist for all the times of interest. The fact that each window must have a limited width need not pose a problem so long as an overlapping sequence of such windows can be found between the initial time and the final time of interest, even if the total range considered, $t_f - t_i$, is much greater than $t_p$.

\subsubsection*{Stochastic inflation within a master equation}
\label{sssec:stochinflmaster}

What has this to do with cosmology? Ref.~\cite{OpenEFT} shows in some detail how the above formalism applies to the physics of extra-Hubble modes during inflation. (See also \cite{OtherNoiseDeriv}.) In this case sector $A$ is taken to be the set of field modes satisfying $k/a \ll H$ with the rest of the modes making up sector $B$. Within a semiclassical calculation write a quantum field, $\Phi$, as $\Phi = \varphi + \phi$ where $\varphi$ is the classical background and $\phi$ the quantum fluctuation, and define $H_\ssA$ and $H_\ssB$ as the terms in $H$ involving only super-Hubble (or only sub-Hubble) modes. In practice for weakly interacting fields we take both to be quadratic in $\phi$. In the interaction picture this corresponds to taking the fields $\phi$ to evolve according to the wave operator defined by the background spacetime. $H_\AB$ contains all parts of $H$ that mix the long- and short-wavelength modes. The correlation time in this picture is of order the Hubble time, $t_c \sim H^{-1}$.

Within this framework the leading evolution of the state of the long-wavelength sector, $\rho_\ssA$, for times $t \gg t_c \sim H^{-1}$ is given by an equation of the form of \pref{coarseschem} where all of the interactions $H_\AB$ are dropped. Consequently $\rho_\ssA$ does not evolve at all in the interaction picture or, equivalently, in the Schr\"odinger picture $\rho_\ssA$ evolves with a `free' Liouville equation that sees only the interactions with the classical background. The functional Schr\"odinger equation as applied to the diagonal elements, $P[\varphi] = \langle \varphi | \rho_\ssA | \varphi \rangle$, of the density matrix (in a field basis in coarse-grained position space) becomes the Fokker-Planck equation of stochastic inflation \cite{StochInf}. For the present purposes what is important is that its solutions can reliably capture the late-time behaviour of extra-Hubble modes precisely because it can be derived as the leading approximation to a master equation analysis (which, after all, is designed precisely for this purpose).

There is also a bonus. Because the neglect of $H_\AB$ means the system is basically free the off-diagonal components, $\langle \varphi | \rho_\ssA | \tilde \varphi \rangle$, do what they must for $\rho_\ssA$ to remain a pure state. This is no longer true once one works to quadratic order in $H_\AB$, however, and \cite{OpenEFT} argues that these instead get driven to zero with time (with the `pointer' basis very generally chosen as the field basis by the extra-Hubble squeezing of states). For a broad class of systems the dimensional estimate given in \cite{OpenEFT} indicates that this decohering of long-wavelength modes happens quickly enough that 50-60 $e$-foldings are likely ample for its completion.

\subsection{Evidence for stochastic resummation}
\label{ssec:resumevidence}

So much for generalities. If a stochastic formulation captures the late-time limit of the master equation for fluctuations in inflationary cosmology, how does this help in practice with the IR secular effects encountered \cite{InfPred} when making precise inflationary predictions?

In the stochastic picture correlation functions are computed using the probability distribution, $P(\varphi,t)$, whose time evolution is predicted using the appropriate Fokker-Planck equation. If late-time solutions of this equation are to capture the results of slowly accumulating IR secular effects, then it should be true that the rate of change of correlators predicted from the Fokker-Planck equation agree with the evolution found for these correlators using standard techniques of quantum field theory on curved space, at least for the IR singular part.

Ref.~\cite{TsamisWoodard} tests this proposal in some detail for the specific case\footnote{ Ref.~\cite{TsamisWoodard} also explores examples involving scalars self-interacting through derivative couplings.} of a massless spectator scalar field in de Sitter, self-interacting through a potential $V = \frac{1}{4!} \, \lambda \phi^4$. They do so by isolating the IR singular, time-dependent part of scalar-field correlators on de Sitter space and computing their rate of change with time. Following \cite{StochInf} they argue the IR fields behave like stochastic variables and show that their evolution is governed by a probability density, $P(\varphi,t)$, that satisfies the appropriate Fokker-Planck equation:\footnote{We argue for corrections to this equation in later sections.}
\be \label{StochEqStandard}
 \partial_t P = \frac{H^3}{8\pi^2} \, \left( \frac{\partial^2 P}{\partial\varphi^2} \right) + \frac{1}{3H} \, \frac{\partial}{\partial \varphi} \left( \frac{\partial V}{\partial \varphi} \; P \right) \,,
\ee
with $V(\varphi) = \frac{1}{4!} \, \lambda \varphi^4$. Since the evolution equation for the IR part of the field agrees over a long time period with the Fokker-Planck equation, it shows that the late-time implications of the IR secular evolution can be obtainable from the Fokker-Planck equation's late-time ({\em i.e.} static) solutions. For instance, on the stochastic side the predicted evolution for $\langle \phi^{2n} \rangle$ in ref.~\cite{TsamisWoodard} is (up to an overall -- potentially IR divergent -- additive constant)
\bea \label{phi2nstoch}
 \langle  \phi^{2n} \rangle_{\rm stoch} &=& (2n-1)!! \left( \frac{H^2}{4\pi^2} \; \ln a \right)^n \left[ 1 - \frac{n(n+1)}{2} \left( \frac{\lambda}{36\pi^2} \; \ln^2 a \right) \right. \\
 && \qquad\qquad\left. + \frac{n}{280} ( 35n^3 + 170n^2 + 225n + 74) \left( \frac{\lambda}{36\pi^2} \; \ln^2 a \right)^2 + \cdots  \right] \,,\nn
\eea
where $a = e^{Ht}$ is the inflationary scale factor, whose presence flags the secular evolution of $\langle \phi^{2n} \rangle$ and the eventual breakdown of the $\lambda$ expansion at late times. This agrees with the IR part of the same quantity computed using quantum field theory on de Sitter space.

This example shows the virtue of the stochastic formulation. Although the logarithms of $a$ imply predictions like \pref{phi2nstoch} must break down for moderately large $t$, the same is not true for \pref{StochEqStandard} which can be used to predict the late time-limit, $\cP(\varphi) = \lim_{t \to \infty} P(\varphi,t)$, usually taken to be the time-independent solutions,
\be \label{latenongauss}
 \cP = C \, \exp \left( - \frac{8\pi^2 V}{3H^4} \right) = C\, \exp \left( - \frac{\pi^2 \lambda \varphi^4}{9H^4} \right) \,,
\ee
where $C$ is a $\varphi$-independent normalization. This shows how the statistics of fluctuations at very late times can be very non-gaussian despite the assumption that fluctuations for each mode are individually gaussian as they pass through horizon exit. The secular evolution is the theory's way of telling us this is possible: small secular perturbations accumulating over long times can build up to produce large effects.

\section{The future is stochastic}
\label{section:future}

We now explore some of the previous section's implications; in particular of the conclusion that late-time evolution of $P(\varphi,t) = \langle \varphi | \rho(t) | \varphi \rangle$ is governed by stochastic evolution, dominated by instantaneously gaussian vacuum fluctuations as each mode passes through horizon exit.

\subsection{Implications of near-gaussian horizon-exit}
\label{section:generalities}

The assumption that modes are gaussian at horizon exit (as would be driven by the Bunch-Davies vacuum \cite{BDV} of a weakly interacting quantum field) is a strong one since the stochastic evolution is then determined by the evolution of the mean and variance.

In general the Fokker-Planck equation for the probability $P(\varphi, t)$ of a gaussian system has the form
\be \label{FPdef}
 \frac{\partial P}{\partial t} = \frac{\partial}{\partial \varphi} \left\{ \cN(\varphi,t) \frac{\partial P}{\partial \varphi} + \cK (\varphi,t) P \right\} = \frac{\partial}{\partial \varphi} \left\{ \frac{\partial}{\partial \varphi} \Bigl[ \cN(\varphi,t) \, P \Bigr] + \cF (\varphi,t) P \right\} \,,
\ee
whose coefficients $\cN(\varphi,t)$ and $\cK(\varphi,t)$ are in general functions of $\varphi$ and time. The last equality defines for later convenience the `force' $\cF := \cK - \partial \cN / \partial \varphi$. At least one derivative must stand on the far left of the right-hand side of (\ref{FPdef}) to ensure the normalization of $P$ is preserved in time.

The functions $\cN(\varphi,t)$ and $\cF(\varphi,t)$ determine the time-evolution of the mean and variance. For instance, an integration by parts shows
\be \label{FPmean}
 \partial_t \langle \phi \rangle = \int \exd \varphi \; \varphi \, \frac{\partial}{\partial \varphi} \left[ \frac{\partial}{\partial \varphi} \Bigl( \cN \, P \Bigr) + \cF \, P \right] = - \langle \cF(\phi) \rangle \,,
\ee
and similarly
\be
 \partial_t \langle \phi^2 \rangle =  \int \exd \varphi \; \varphi^2 \, \frac{\partial}{\partial \varphi} \left[ \frac{\partial}{\partial \varphi} \Bigl( \cN \, P \Bigr) + \cF \, P \right] = 2 \langle \cN(\phi) - \phi\, \cF(\phi) \rangle \,.
\ee
The evolution of the variance is therefore given by
\be \label{FPvar}
 \partial_t \Bigl( \langle \phi^2 \rangle - \langle \phi \rangle^2 \Bigr) = 2 \langle \cN(\phi) \rangle - 2 \Bigl[ \langle \phi\, \cF(\phi) \rangle - \langle \phi \rangle \, \langle \cF(\phi) \rangle \Bigr] \,,
\ee
and so receives contributions both from the `noise' $\cN$ and the fluctuations of the `work' done by the force $\cF$. 

In principle $\cN$ and $\cF$ can be computed given the wave-functional --- in de Sitter space the Bunch-Davies vacuum, for example --- describing the state of each mode. However when this is done explicit calculations (see Appendix \ref{App:fluccalc} and \cite{OpenEFT}) give coefficients, $\cN = \cN(t)$ and $\cF = \cF(t)$, that are functions of time only. This is not so useful for forecasting late-time evolution because these functions of time are themselves only known perturbatively, and so are plagued by secularly growing terms.

More useful is if $\cN$ and $\cF$ are directly related to $\varphi$, if this can be done in a way that holds instantaneously for all $t$, because then the Fokker-Planck equation integrates to give nontrivial information about the late-time evolution. In general this is not possible, since $\partial_t \langle \phi \rangle$ is usually not uniquely determined by $\langle \phi \rangle$. It {\em can} be possible in certain circumstances, however, and when this is possible the direct connection between $\partial_t \langle \phi \rangle$ and $\langle \phi \rangle$ allows $\cF$ and $\cN$ to be determined as functions of $\varphi$.

An important example of this type is slow roll, for which the long-wavelength modes of a quasi-free field satisfy
\be \label{slowrollstandard}
 0 = \partial^2_t \langle \phi \rangle + 3H \partial_t \langle \phi \rangle + m^2 \langle \phi \rangle \approx  3H \partial_t \langle \phi \rangle + m^2 \langle \phi \rangle \,,
\ee
where the approximate equality neglects $\partial_t^2 \langle \phi \rangle$ relative to $H \partial_t \langle \phi \rangle$. Comparing with \pref{FPmean} gives the usual result
\be \label{Fvsm}
 \cF(\varphi) \approx  \left( \frac{m^2}{3H} \right) \varphi \,,
\ee
(some corrections to which are described below). The Starobinsky result of the previous section corresponds to assuming the noise is dominated by its massless limit, so $\cN = H^3/8\pi^2$, and generalizing \pref{Fvsm} to the case where the force is a slowly varying function of $\langle \phi \rangle$, given in terms of the scalar potential, $V$, by $\cF(\varphi) = V'(\varphi)/3H$.

But the generality of the argument is not restricted to these choices, and later sections explore how they can differ for a few other cosmological scenarios. Before doing so we first digress to give the general forecast for the late-time distribution, $\lim_{t \to \infty} P(\varphi, t)$, as a function of the coefficients $\cN(\varphi)$ and $\cF(\varphi)$.

\subsection*{Late-time limit for $P$}

The time evolution of the Fokker-Planck equation generally describes a slow relaxation towards a static solution, $\lim_{t \to \infty} P(\varphi, t) = \cP(\varphi)$, so it is the static solutions that are of most interest at late times. As the previous section makes clear, this late-time form need not remain gaussian due to the accumulation over long times of locally small effects, and expresses the late-time forecast implied by IR secular effects.

The expression for $\cP(\varphi)$ can be found explicitly by integrating the Fokker-Planck equation when $\cN$ and $\cF$ are functions of $\varphi$ only. Demanding $\partial_t \cP = 0$ gives
\be
 \frac{\partial}{\partial \varphi} \left\{ \frac{\partial}{\partial \varphi} \Bigl[ \cN(\varphi) \, \cP \Bigr] + \cF(\varphi) \, \cP \right\} = 0 \,,
\ee
whose general solution is
\be \label{GenLateFP}
 \cP(\varphi) = \left[ \frac{k_1 \varphi + k_2}{\cN(\varphi)} \right] \exp \left[ - \int \exd \varphi \; \frac{\cF(\varphi)}{\cN(\varphi)} \right] \,,
\ee
where $k_1$ and $k_2$ are integration constants. In the special case where $\cN = H^3/8\pi^2$ is $\varphi$-independent and $\cF = V'/3H$ this reduces to
\be
 \cP \to \frac{8 \pi^2}{H^3} \, (k_1 \varphi + k_2) \exp \left( - \frac{8 \pi^2 V}{3H^4} \right) \,,
\ee
agreeing with the standard Starobinsky result (for which $k_1 = 0$ is usually chosen and $k_2$ is fixed by normalization). Notice that a prerequisite for $\cP(\varphi)$ to be an IR safe quantity is that both $\cN$ and $\cF$ must also be IR safe.

\subsection{Masses, sound speeds,  and non-de Sitter expansion}
\label{section:variationsI}

We next explore these arguments in more detail by extending eq.~\pref{StochEqStandard} away from de Sitter space, in a simple enough way to allow explicit exact solutions but also complicated enough to illustrate the connection between IR divergences and secular behaviour while keeping the calculations relatively simple. To avoid single-field no-go arguments we work within a multiple-field framework, but for simplicity restrict ourselves initially to the case where any additional scalars are spectators inasmuch as they play no direct role in the rate of inflationary expansion. In this section we assume a spectator scalar mass for which $m^2/H^2$ is time-independent, and return in the next section to the broader extension to self-interactions and field-dependent masses. We adjust the following three dials in what follows:\footnote{A fourth dial -- the possibility the scalar has a non-standard dispersion relation, such as in ghost inflation \cite{ArkaniHamed:2003uz} -- is explored in Appendix \ref{app:ghostinf}.}

\medskip\noindent{\em 1. Power-law evolution:}

\smallskip\noindent We consider power-law inflating spacetimes,
\be
 a(t) = a_0 \left( \frac{t}{t_0} \right)^p
\ee
for constant $p > 1$. Unlike near de Sitter spacetimes the Hubble and first slow-roll parameters depend on time as
\be \label{powerlawdef}
 H(t) = \frac{p}{t} = H_0 \left( \frac{a_0}{a} \right)^\epsilon  \quad \hbox{where} \quad \epsilon := -\frac{\dot H}{H^2} = \frac{1}{p} \,,
\ee
and so $H$ varies with time while $\epsilon$ is constant. We demand $p > 1$ to ensure the spacetime expansion accelerates (so that modes exit the Hubble scale as usual as time evolves). The geometry reduces to the exponential expansion of de Sitter space in the limit $p \to \infty$ with $H_0 = p/t_0$ fixed, though we do not necessarily require $p \gg 1$ in much of what follows.

\medskip\noindent{\em 2. Nonzero masses:}

\smallskip\noindent
We track the stochastic description of vacuum fluctuations for a free spectator scalar field with small nonzero mass, and unlike for \pref{StochEqStandard} we seek explicit expressions for how {\em both} the noise and drift vary with nonzero $m$. We allow $m^2$ to be time-dependent but (for simplicity) to do so in such a way that $m^2/H^2 \ll 1$ is time-independent.  We return below to a discussion of some implications of weak interactions, including the possibility of having a field-dependent mass.\footnote{Masses could be time-dependent without depending on the field $\phi$ itself due to couplings with other fields (such as the inflaton) which we do not consider here.}

\medskip\noindent{\em 3. Varying sound speed:}

\smallskip\noindent
Finally, we track the implications of a small time-dependent speed of sound parameterized by
\be \label{csdef}
 c_s = c_0 \, (a/a_0)^s \,,
\ee
with constant $s$ and $c_0 \ll 1$ chosen so that $c_s$ remains smaller than unity for the entire time interval of interest. Cosmological models with varying sound speed like this are studied, for example, in \cite{Varyingcs}.

\subsubsection*{Quantum fluctuations}

The first step is to compute how vacuum fluctuations cause the mean and variance to vary during horizon exit. As discussed above this leads to predictions for $\cN(t)$ and $\cF(t)$ that are functions of $t$ rather than $\varphi$, whose explicit form --- for nearly free fields and the above assumptions concerning $c_s$ and $a(t)$ --- can be computed as easily as for de Sitter space. 

The solutions are found by constructing explicitly the vacuum wave-functional (in Schr\"odinger picture), with results summarized here (details of this calculation alattre given in Appendix \ref{App:fluccalc}, with generalizations to other dispersion relations given in Appendix \ref{app:ghostinf}). This in turn can be computed from scalar-field mode functions satisfying the Klein-Gordon equation in the spacetime of interest. For a spatially flat FRW metric,
\be
  \exd s^2 = - \exd t^2 + a^2(t) \, \exd \vec{x} \cdot \exd \vec{x} \,,
\ee
the modes, $u_k(t) \, e^{i \vec{k} \cdot \vec{x}}$, are labelled by co-moving momentum, $\vec k$, and satisfy
\be
   \ddot u_k + 3 H \dot u_k + \left[ \left( \frac{c_s k}{a} \right)^2 + m^2 \right] u_k = 0 \,,
\ee
with dots denoting derivatives with respect to $t$. The solutions, $u_k$, determine the kernels, $\alpha_k$ and $\beta_k$, appearing in the corresponding ground-state wavefunctional for the scalar field, given (in the Schr\"odinger picture) by $\Psi = \prod_k \Psi_k$ with
\be \label{defpsik}
  \Psi_k[\varphi_k] = C_k \; \exp \left[ - a^3 \, \left( \alpha_k \, \varphi_k \varphi_{-k} + \beta_k \, \varphi_k \right) \right] \,.
\ee

Explicitly, the functional Schr\"odinger equation relates $\alpha_k$ and $\beta_k$ to the knobs we are free to dial: our constant choices for $m^2/H^2$, $s$ and $\epsilon$ --- with the latter two arising in the choices \pref{powerlawdef} and \pref{csdef}. As shown in Appendix \ref{App:fluccalc}, the relationship between these variables is given in terms of the mode functions by
\be
 \alpha_k = -i \left( \frac{\dot u_k}{u_k} \right)\,,
\ee
and
\be
 \beta_k = \bar{\beta}_0 \, \delta_{k0} \, \left( \frac{a_0}{a(t)} \right)^3   \exp \left[ -i \,\int^t_0 \, \exd \tau \,\alpha_0(\tau)  \right] 
 =  \bar{\beta}_0 \, \delta_{k0} \, \left( \frac{a_0}{a(t)} \right)^3   \frac{u_0(0)}{u_0(t)} \,,
\ee
where $\bar\beta_0 = \beta_0(t=0)$ is an integration constant that turns out to be fixed by the initial value of the expectation value of the field, $\langle \phi \rangle(t=0)$. For nonzero $\vec k$ the mode function that properly extrapolates from the adiabatic vacuum in the limit $c_s k/a H \gg 1$ turns out to be
\be \label{modeform}
  u_k(t) = C_k \, y^q(a,k) \, H_\nu^{(2)}[y(a,k)] \,,
\ee
with independent variable, $y$, given in terms of $a$ and $k$ by
\be
   y(a,k) := \frac{1}{(1-s-\epsilon)}\left( \frac{c_s\,k}{a H} \right) = \frac{1}{(1-s-\epsilon)}\left( \frac{c_0\,k}{a_0 H_0} \right) \left( \frac{a_0}{a} 
   \right)^{1-s-\epsilon}   \,,
\ee 
and the power $q$ given by
\be
 q = \frac{3-\epsilon}{2\, (1-s-\epsilon)} \,.
\ee
Here $H_\nu^{(2)}(y)$ is the Hankel function of the second kind, of order
\be \label{eqfnut}
  \nu^2 = \frac{1}{(1-s-\epsilon)^2}\left[  \frac{(3-\epsilon)^2}{4}-\frac{m^2}{H^2}\right] = q^2 \left[  1 -\frac{4m^2}{(3-\epsilon)^2H^2} \right] \,,
\ee
Notice that these expressions reduce to the standard ones for a massive field on de Sitter space (constant $H$) when both $\epsilon \to 0$ and $s \to 0$. 

\subsubsection*{Evolution of the mean}

To compute the noise and drift (and to explore the connection between IR singularities and secular late-time evolution) we first require expressions for the rate of change of the mean and variance of the quantum field, in position space and coarse-grained over an extra-Hubble volume,
\be \label{cgfield}
 \phi_\cS(\vec r,t) =   \int \frac{\exd^3 k}{(2\pi)^3}\, {\cal S}\left[y(k) \right] \,
 \phi_k  \, e^{i \vec{k} \cdot \vec{r} } \,,
\ee
where ${\cal S}$ is the window function that coarse-grains over sub-Hubble modes. The details of $\cS$ are not important in most of what follows, with the main results relying only on the limiting properties
\be \label{Slimst}
  {\cal S}\left(y\right)  \to  1 \quad \hbox{for $y\ll1$} \qquad \hbox{and} \qquad
  {\cal S}\left(y\right) \to  0 \quad \hbox{for $y\gg1$} \,.
\ee

Since the quantum state is gaussian all fluctuations in can be computed explicitly as functions of the known quantities $\alpha_k$ and $\beta_k$ (see Appendix \ref{App:fluccalc} for details). The mean of the field is\footnote{We assume translation-invariant backgrounds for which only the $k=0$ mode contributes to $\langle \phi_\cS \rangle = \langle \phi \rangle$.}
\be
  \langle \phi_\cS \rangle = \langle \phi \rangle = \frac{\beta_0+\beta_0^*}{2\left(\alpha_0+\alpha_0^* \right)} \,,
\ee
and so its time evolution is given by the equations of motion for $\alpha_0$ and $\beta_0$ as
\be \label{phidot}
  \partial_t\,\langle \phi_\cS \rangle
  = \partial_t\,\langle \phi \rangle
  = i \left( \frac{\alpha_0 \beta_0^* - \beta_0 \alpha_0^*}{\alpha_0 + \alpha_0^*} \right) \,,
\ee
which uses translation invariance and the limit $\cS \to 1$ as $k \to 0$. This expresses the standard relation between $\partial_t \langle \phi \rangle$ and the canonical momentum, $\langle \Pi \rangle = a^3 \partial_t \, \langle \phi \rangle$. 

Differentiating once more leads to Ehrenfest's theorem for this system,
\be
 \partial_t^2 \langle \phi_\cS \rangle + 3H \partial_t \langle \phi_\cS \rangle + m^2   \langle \phi_\cS \rangle = 0 \,,
\ee
stating that the mean satisfies the classical equations of motion. This has simple power-law solutions, $\langle \phi_\cS \rangle \propto (a_0/a)^{r_\pm}$ with 
\be
 r_\pm = \frac{3-\epsilon}2 \left[ 1 \pm \sqrt{1 - \frac{4 m^2}{(3-\epsilon)^2 H^2}} \right] = ( q \pm \nu) (1 - s - \epsilon) \,,
\ee
which for small mass becomes
\be
 r_+ \simeq 3-\epsilon \qquad \hbox{and} \qquad 
 r_- \simeq \frac{m^2}{(3-\epsilon)H^2} \,.
\ee
Of these $r_-$ describes the more slowly decaying\footnote{This solution must become static as $m^2 \to 0$ because it must cross over to slowly grow once $m^2 < 0$, reflecting the tachyonic instability.} solution (when $0 \le m^2 \ll H^2$) that typically dominates at late times. Notice that when it does $\partial_t \langle \phi_\cS \rangle$ is directly related to $\langle \phi_\cS \rangle$ by the slow-roll condition
\be \label{SlowRollRel}
  \partial_t \langle \phi_\cS \rangle = -r_- H \langle \phi_\cS \rangle = \left[ -\frac{m^2}{(3-\epsilon)H} + \cdots \right] \langle \phi_\cS \rangle \,,
\ee
which reduces to the approximate equality of \pref{slowrollstandard} only to lowest order in $m^2/H^2$.

\subsubsection*{Evolution of the variance }

For each mode the variance about the mean similarly is
\be
  \langle \phi_k  \phi_k^* \rangle  = \frac{1}{a^3\left(\alpha_k+\alpha_k^*\right)}  = |u_k|^2 \,,
\ee
so the coarse-grained position-space two-point function becomes
\be \label{varexpt}
  \left \langle  (\phi_\cS - \langle \phi_\cS \rangle)^2 \right\rangle  =  \int \frac{\exd^3 k}{(2\pi)^3}\,|u_k\, {\cal S}|^2 \,.
\ee
Its rate of change then evaluates (see Appendix \ref{App:fluccalc}) to
\bea 
  \partial_t \left \langle  (\phi_\cS - \langle \phi_\cS \rangle)^2 \right\rangle &=&  H (1 - s - \epsilon) \left[ \frac{1}{2 \pi^2} \lim_{k \to 0} 
  \left( k^3 |   u_k|^2 \right) + (3-2q) \int \frac{\exd^3 k}{(2\pi)^3} \, |u_k\,{\cal S}|^2 \right] \ \nn \\
   &=&   (1 - s - \epsilon) \frac{H}{2 \pi^2} \lim_{k \to 0} \left( k^3 |u_k|^2 \right)  \\
   && \qquad\qquad\qquad\qquad\qquad  - (3s+2\epsilon)H\left \langle  (\phi_\cS - \langle 
   \phi_\cS   \rangle)^2 \right\rangle  \,.\nn
 \label{eqf2t}
\eea

What is important about \pref{eqf2t} is that it holds regardless of the details of the mode functions, $u_k$, since it relies only on the property that the variables $a$ and $k$ appear in $k^{2q} |u_k \cS|^2$ exclusively through the combination $y(a,k)$. This implies that the time-evolution of the variance for a wide variety of states can be found by regarding \pref{eqf2t} as a differential equation for the quantity $Y(t) := \langle (\phi_\cS - \langle \phi_\cS \rangle)^2 \rangle$ and integrating. 

Appendix \ref{App:fluccalc} shows the general solution (for constant $s$, $\epsilon$ and $m^2/H^2$) is given by
\bea \label{Fasoln2}
 Y(a) &=& \left\{ \frac{c_0^3 Y_0}{H_0^2} - \frac{K(\nu)}{(3-2\nu)(1-s-\epsilon)} \left[ \left( \frac{a_0}{a} \right)^{(3-2\nu)(1-s-\epsilon)} -1 \right] \right\} \frac{H^2(a)}{c_s^3(a)}  \\
 &=& \left[ Y_0 + \frac{K(\nu) H_0^2/c_0^3}{(3-2\nu)(1-s-\epsilon)} \right] \left( \frac{a_0}{a} \right)^{2\epsilon+3s} - \frac{K(\nu) H_0^2/c_0^3}{(3-2\nu)(1-s-\epsilon)} \left( \frac{a_0}{a} \right)^{2(q-\nu)(1-s-\epsilon)} \,, \nn
\eea
where $Y_0 = Y(t=t_0)$ denotes the initial variance and
\be \label{Kdef}
 K(\nu) := \frac{ |2^\nu \Gamma(\nu) (1 - s - \epsilon)^\nu |^2}{(2 \pi)^3}  \lim_{\mu \to 0} \left( \frac{\mu c_0}{a_0 H_0} \right)^{3-2\nu} \,,
\ee
comes from evaluating $k^3|u_k|^2$ using the specific choice for $u_k$ given in eq.~\pref{modeform}. 

The second of eqs.~\pref{Fasoln2} shows that $Y(a)$ generically decays with time because the two powers are non-negative for $0 \le s, \epsilon < 1$ and $0 \le m^2 \le (3-\epsilon)^2 H^2/4$, with
\be
 2(q-\nu)(1-s-\epsilon) = (3-\epsilon) \left[ 1 - \sqrt{1 - \frac{4m^2}{(3-\epsilon)^2 H^2}} \,\right] \approx \frac{2m^2}{(3-\epsilon)H^2} + \cdots \,,
\ee
while $Y$ asymptotes to a constant in the special case of a massless field. The first of eqs.~\pref{Fasoln2} is the more useful when taking the $\nu \to \frac32$ limit, giving
\be
  Y(a) \to \left[ \frac{c_0^3 Y_0}{H_0^2} + K_{3/2} \ln \left( \frac{a}{a_0} \right) \right] \frac{H^2(a)}{c_s^3(a)}   \qquad \hbox{(if $\nu \to \frac32$)}\,,
\ee
where 
\be
  K_{3/2} := K(\nu \to 3/2) = \frac{(1-s-\epsilon)^3}{(2\pi)^2}  \,. 
\ee
This agrees with standard results for massless fields in de Sitter space in the limit $\epsilon \to 0$ (such as the leading term in eq.~\pref{phi2nstoch} if $c_s=n=1$ and $Y_0$ is also chosen to vanish), in which case $\ln(a/a_0) \to H_0(t-t_0)$ shows the usual linear growth with $t$. 

The above also illustrates the connection between uncontrolled secular growth and IR singularities, as follows. On one hand eq.~\pref{Kdef} shows that $K(\nu)$ diverges in the IR (as $\mu \to 0$) if and only if $\nu > \frac32$. On the other hand \pref{Fasoln2} shows (for the parameter range of interest) that $Y(a)$ also grows without bound relative to the natural benchmark $H^2/c_s^3$ if and only if $\nu > \frac32$. Because $H^2/c_s^3$ generically falls as $a$ grows this relative growth of $c_s^3 Y/H^2$ at best ensures $Y$ remains constant in time, such as in the massless limit for which $\nu \to q$ and \pref{Fasoln2} reduces to
\be
   Y(a) \to \left[ \frac{c_0^3 Y_0}{H_0^2} - \frac{K_\IR}{2\epsilon + 3s} \right] \frac{H^2(a)}{c_s^3(a)} + \frac{K_\IR H_0^2/c_0^3}{2\epsilon + 3s}
   \qquad \hbox{(if $m^2 \to 0$)} \,,
\ee
where 
\be
 K_\IR := K(m^2 \to 0) = \frac{ |2^q \Gamma(q) (1 - s - \epsilon)^q |^2}{(2 \pi)^3}  \lim_{\mu \to 0} \left( \frac{\mu c_0}{a_0 H_0} \right)^{3-2q} 
\ee
is singular as $\mu \to 0$ for small $s$ and $\epsilon$ since $q  > \frac32$. Notice that this singularity would appear as a logarithmic IR divergence
\be
 K_\IR = K_{3/2} \left\{ 1 + (3-2q) \left[ \lim_{\mu \to 0} \ln \left( \frac{c_0 \mu}{a_0 H_0} \right) + \hbox{finite} \right] + \cO \left[(3-2q)^2 \right] \right\} \,,
\ee
in an expansion about de Sitter space.

\subsubsection*{Noise \& Drift}

We next return to the Fokker-Planck equation and use these results to read off the noise and drift functions, $\cN(\varphi)$ and $\cF(\varphi)$, and what is remarkable is that these {\em always} remain IR finite even when the variance diverges. To see how this works we demand $\cN$ and $\cF$ in eqs.~\pref{FPmean} and \pref{FPvar} reproduce the above expressions for the variation of the mean and variance computed from the Schr\"odinger-picture wave-functional. 

Since the wave-functional gives results directly as functions of time only, it is tempting (but not that useful) to seek $\cF = \cF(t)$ and $\cN = \cN(t)$ that also depend only on time, in which case we would find
\be
  \cF(t) = - \partial_t\,\langle \phi_\cS \rangle
  = -i \left( \frac{\alpha_0 \beta_0^* - \beta_0 \alpha_0^*}{\alpha_0 + \alpha_0^*} \right) \,,
\ee
and 
\bea \label{Noft}
 \cN(t) &=& \frac12 \, 
  \partial_t \left \langle  (\phi_\cS - \langle \phi_\cS \rangle)^2 \right\rangle \nn\\
  &=&  (1 - s - \epsilon) \frac{H}{4 \pi^2} \lim_{k \to 0} \left( k^3 |u_k|^2 \right) - \frac{H}2 (3s+2\epsilon)  \int \frac{\exd^3 k}{(2\pi)^3}\,|u_k\, {\cal S}|^2 \,.
\eea
These are not that useful because to know them as functions of $t$ requires already knowing the late-time behaviour, so they do not add new capabilities to resum secular evolution. 

Nor do $\cF(t)$ and $\cN(t)$ have better IR behaviour than does the rate of change of the mean and the variance. In particular, to uncover the IR behaviour we use $u_k \propto k^{-\nu}$ to see that $k^3 |u_k|^2 \to A k^w$ for small $k$, with power
\bea
    w = 3 - 2\nu &=& (3-2q) + 2(q-\nu) \nn\\
    &=& - \frac{3s+2\epsilon}{1 - s - \epsilon} + \frac{3-\epsilon}{1 - s - \epsilon} \left[ 1 - \sqrt{1 - \frac{4m^2/H^2}{(3-\epsilon)^2}} \;\right] \nn\\
    &\approx&   - \frac{3s+2\epsilon}{1 - s - \epsilon} + \frac{2m^2/H^2}{(3-\epsilon)(1 - s - \epsilon)} + \cO(m^4/H^4) \,,
\eea
that can be nonpositive in the regime $0 \le m^2/H^2 \le 3s + 2 \epsilon$. Consequently the contribution from $k \to 0$ to the right-hand side of \pref{Noft} becomes
\bea
 \cN_\IR &=&  \frac{AH}{4 \pi^2}\lim_{k\to 0} \left[ (1 - s - \epsilon) k^w  -  (3s+2\epsilon)  \int_k \exd u u^{w-1} \right] \nn\\
 &=& (1 - s - \epsilon) \frac{AH}{4 \pi^2}\lim_{k\to 0} \left[ k^w  +  \frac{1}{w} (2q-3)  k^w \right] + (\hbox{finite}) \nn\\
  &=& (1 - s - \epsilon) \left( \frac{\nu - q}{2\nu -3} \right) \frac{AH}{2 \pi^2}\lim_{k\to 0} k^w + (\hbox{finite})
\eea
and so diverges if $w \le 0$ and $m^2 \ne 0$. 

\bigskip\noindent{\em $\varphi$-dependence and IR Finiteness}

\medskip\noindent
Following the general discussion of the earlier sections, we expect these IR singularities to be better described in situations where $\cF$ and $\cN$ are computed as functions of $\varphi$ rather than $t$, since in this case the late-$t$ limit can be found by integrating the FP equation rather than through direct calculation in an expansion about free fields. We now show that the noise also becomes IR finite when this is done.

Having $\cN = \cN(\varphi)$ and $\cF = \cF(\varphi)$ is in general not possible since it requires the rate of change of the mean and variance to be dictated purely by an instantaneous average over $\varphi$. They {\em can} be so related in the special case of slow evolution, however, since in the slow-roll regime eq.~\pref{SlowRollRel} holds, implying
\be \label{SlowRollAgain}
  \partial_t \langle \phi_\cS \rangle \simeq (\nu-q)(1-s-\epsilon) H \langle \phi_\cS \rangle \simeq \left[- \frac{m^2}{(3-\epsilon)H} + \cdots \right]  \langle \phi_\cS \rangle \,,
\ee
which generalizes the usual de Sitter slow-roll relation. We emphasize that because $\phi$ is a spectator field (and not the inflaton) this slow-roll condition need {\em not} also require the metric rolls equally slowly --- {\em ie} $\epsilon$ can be much larger than $m^2/H^2$. 

Comparing \pref{SlowRollAgain} with \pref{FPmean} --- and using that $\cF(\varphi)$ is linear in $\varphi$ for gaussian systems ---gives the generalization of the usual Starobinsky result
\be \label{Fvarphi}
 \cF(\varphi) = (q-\nu)(1-s-\epsilon) H \varphi  \simeq \left[ \frac{m^2}{(3-\epsilon)H} + \cdots \right]  \varphi  \,,
\ee
Using this in \pref{FPvar} and comparing with \pref{eqf2t} then gives
\bea \label{NoftIRsafe}
   \cN   &=&  (1 - s - \epsilon) \frac{H}{4 \pi^2} \lim_{k \to 0} \left( k^3 |u_k|^2 \right) +H \left[ (q-\nu)(1-s-\epsilon) - \frac{3s+2\epsilon}{2} 
    \right] \left \langle  (\phi_\cS - \langle \phi_\cS \rangle)^2 \right\rangle \nn\\
    &=& (1 - s - \epsilon) \frac{H}{4 \pi^2} \left\{ \lim_{k \to 0} \left( k^3 |u_k|^2 \right) + \Bigl[ 2(q-\nu) - (2q - 3) \Bigr]
    \int_0^\infty \exd k \; k^2 \, |u_k\, {\cal S}|^2 \right\} \,, \nn\\
    &=& (1 - s - \epsilon) \frac{H}{4 \pi^2} \left\{ \lim_{k \to 0} \left( k^3 |u_k|^2 \right) - (2 \nu - 3)  \int_0^\infty \exd k \; k^2 \, |u_k\, {\cal S}|^2 \right\} \,,
\eea
where for gaussian systems we take $\cN$ to be $\varphi$-independent and so identify $\cN = \langle \cN \rangle$. Notice that the term proportional to $(\nu - q)$ comes from the fluctuation of the drift force, and is precisely what is required to make the result for $\cN$ IR finite (as also found in \cite{OpenEFT}). Notice that, in general, the noise depends on the window function used to build our coarse-grained variable.  See also \cite{Cho:2015pwa,Prokopec:2015owa} for recent studies of stochastic inflation in set-ups with  departures from a pure de Sitter geometry, and 
  \cite{Vennin:2015hra} for a recent paper discussing stochastic corrections to inflationary observables.

For numerical purposes it is useful to express $u_k$ in terms of Hankel functions and make the cancellation of IR divergences more explicit by adding and subtracting the appropriate multiple of $\partial_y \left[ y^{3-2\nu} |\cS|^2 \right]$ to the integrand to get 
\be \label{Nvsnu}
   \cN  = \frac{H^3}{8\pi^2 c_s^3} \; \cR_\cS(\nu) \,,
\ee
where
\bea
 \cR_\cS(\nu) &:=& \frac{\pi}{2} \int_0^\infty \exd y \left\{ (3-2\nu) \left[ y^2 \Bigl| H^{(2)}_\nu(y) \Bigr|^2 - |C(\nu)|^2 y^{2-2\nu} \right] |\cS|^2 \right. \\
 && \qquad\qquad\qquad\qquad\qquad\qquad\qquad\qquad\qquad \left. \phantom{\frac12} - |C(\nu)|^2 y^{3-2\nu} \partial_y |\cS|^2 \right\} \,,\nn
\eea
with $C(\nu) := i2^\nu \Gamma(\nu)/\pi$ the coefficient arising in the asymptotic expansion $H^{(2)}(y) \simeq C(\nu) y^{-\nu}$ for small $y$. The virtue of this expression is its manifest convergence as $y\to0$. This is ensured by the cancellation of the leading small-$y$ behaviour between the terms within the square bracket, while the last term is finite because $\partial_y |\cS|^2$ has support only within a region near $y = 1$.  Convergence at $y \to \infty$ is ensured by the falloff $|\cS|^2 \to 0$.

Eq.~\pref{Nvsnu} also emphasizes how $ \cN $ depends on $s$, $\epsilon$ and $m^2/H^2$ only through $\nu$ and direct evaluation shows that $\cR_\cS(\nu = 3/2) = 1$ in agreement with the standard result when $m^2 = s = \epsilon = 0$. Expanding about $\nu = \frac32$ gives the following leading dependence of $\cN$ on the parameters $s$, $\epsilon$ and $m^2/H^2$. 
\be 
  \cR_\cS \simeq 1 +  (3-2\nu) \int_0^\infty \exd y \left[ y + \left(\psi(3/2) + \ln \frac{y}{2} \right) \partial_y \right] |\cS|^2 + \cO\left[ (3-2\nu)^2 \right] \,,
\ee
with $\psi(\nu) := \partial_\nu \ln \Gamma(\nu)$ and
\bea
 3-2\nu &=& \frac{1}{1-s-\epsilon} \left[ - 3s -2\epsilon + (3-\epsilon) \left( 1 - \sqrt{1 - \frac{4 m^2/H^2}{(3-\epsilon)^2}} \right) \right] \nn\\
 &\simeq& \frac{1}{1-s-\epsilon} \left[ - 3s -2\epsilon + \frac{2 m^2}{(3-\epsilon) H^2} + \cO \left( m^4/H^4 \right) \right] \,.
\eea
For the special case where $|\cS|^2 = \Theta(1-y)$ is a step function the integral evaluates to give $\cR_\cS \simeq  1 + \left[\frac92 + 3 \ln 2 + \gamma \right](3-2\nu) \simeq 1 + 7.157 (3-2\nu)$, where $\gamma = - \psi(1) = 0.5772...$ is the Euler-Mascheroni constant, showing how positive $m^2$ acts to increase the noise while positive $s$ and $\epsilon$ decrease it.

\subsection{Late-time limit}
\label{ssec:compother}

Using expressions \pref{Fvarphi} and \pref{Nvsnu} for $\cF$ and $\cN$ the late-time form, \pref{GenLateFP}, for the probability distribution, $\cP$, obtained from the Fokker-Planck equation finally gives
\be \label{GenLateFPnew}
 \cP(\varphi) = \left[ \frac{k_2}{\cN(\nu)} \right] \exp \left[ - \frac{1}{\cN(\nu)} \int \exd \varphi \; \cF(\varphi) \right] = \sqrt{\frac{\alpha}{2\pi H^2}} \; \exp \left[ - \frac{\alpha(\nu)}{2} \left( \frac{\varphi}{H} \right)^2  \right] \,,
\ee
with
\be \label{GenLateFPnew2}
 \alpha(\nu) := \frac{(q-\nu)(1-s-\epsilon)H^3}{\cN(\nu)}  = \frac{8\pi^2 c_s^3 \;(q-\nu)(1-s-\epsilon)}{ \cR_{\cS}(\nu) } \,.
\ee
where we take $k_1=0$ and choose $k_2$ to normalize the result over the interval $(- \infty, \infty)$. This computes the IR-finite corrections to the late-time distributions that arise for free scalar fields as functions of $s$, $\epsilon$ and $m^2/H^2 \ll 1$. These are seen to preserve the gaussian nature of the fluctuations but modify their variance. The exception to this statement is the case $m = 0$ for which $\nu = q$ and so $\cP(\varphi)$ becomes uniform for all $s$ and $\epsilon$. 

In particular, these allow an expression to be derived for the late-time expectation of the extra-Hubble part of the energy density and so to assess whether or not the secular accumulation of IR effects during inflation gives rise to a large gravitational back-reaction. This is most easily done by switching briefly to Heisenberg representation, for which the slow-roll condition holds as an operator statement: $\dot \phi_\cS \simeq (\nu-q)(1 - s- \epsilon) \, H \phi_\cS$. In this case the late-time expectation is $\langle \dot \phi_\cS^2 \rangle_\infty \simeq (\nu-q)^2(1 - s- \epsilon)^2 \, H^2 \langle \phi_\cS^2 \rangle_\infty$, where the above derivation shows that the late-time two-point function resums to the following IR-finite value,
\be
 \langle \phi_\cS^2 \rangle_\infty = \int_{-\infty}^\infty \exd \varphi \; \varphi^2 \cP(\varphi) 
 = \frac{H^2}{\alpha(\nu)} \,.
\ee

Combining these, the late-time expectation of the extra-Hubble part of the energy density becomes
\bea
 \frac12 \left\langle \dot\phi_\cS^2 + m^2 \phi_\cS^2 \right\rangle_\infty &=& \frac{H^4 }{16\pi^2 c_s^3}  \left[(q-\nu)(1 - s- \epsilon) + \frac{m^2/H^2}{(q-\nu)(1 - s- \epsilon) } \right]\, \cR_\cS(\nu) \nn\\
 &=& \frac{ H^4}{16\pi^2 c_s^3 } \,(3-\epsilon) \cR_\cS(\nu) \,,
\eea
which uses $m^2/H^2 = (1 - s - \epsilon)^2(q^2- \nu^2)$. Among other things this shows that the secular accumulation of IR effects during inflation does {\em not} give rise to a large gravitational back-reaction, at least in this instance.

\subsection{Comparison with other techniques}
\label{ssec:compother}

We next compare the above results with several existing calculations: the large-$N$ limit of $N$ self-interacting scalar fields and the dynamical renormalization group (DRG). 

\subsubsection*{Comparison with $\lambda \phi^4$ at large $N$}
\label{ssec:compphi4}

We wish to use our results to probe a limit of interacting scalar fields in order to test their success by comparison with other calculations. One such an application is to $N$ self-interacting canonical scalar fields, $\Phi$, coupled through a scalar potential 
\be
 V = \frac{\lambda}{4!} \, \left( \Phi \cdot \Phi \right)^2 \,,
\ee
in the limit $\lambda \to 0$ and $N \to \infty$ with $g = \lambda N$ held fixed. (See also \cite{LargeNInf} for inflationary calculations for the large-$N$ model.) As summarized in Appendix \ref{app:largeN} at leading order in the $1/N$ expansion this is described by $N$ free scalar fields with mass 
\be
 \frac{m_\phi^2}{H^2} = \frac{\sqrt{g}}{4\pi} \,,
\ee
and because this is both time-independent and small its late-time limit can be described by the results found above. 

In particular, the late-time probability distribution is IR safe and given by \pref{GenLateFPnew} with
\be \label{GenLateLargeN2}
 \alpha(\nu) = 4\pi^2 c_s^3 (3-\epsilon) \cR_{\cS}(\nu) \left[ 1 - \sqrt{1 - \frac{ \sqrt{g}}{\pi(3-\epsilon)^2}} \;\right] 
 \simeq \frac{2\pi  \sqrt{g} \; c_s^3}{3-\epsilon} \; \cR_{\cS}(\nu) \,,
\ee
and $(\nu/q)^2 = 1 - \sqrt{g}/[\pi(3-\epsilon)^2]$. Notice that although this expression relies on the neglect of $1/N$ it does not also require dropping subdominant powers of $\sqrt{g}$. Furthermore, it is clearly nonperturbative in $g$, as might be expected for resummed contributions.

\subsubsection*{$N=1$ and the Hartree approximation}
\label{ssec:hartree}

These same arguments do not straightforwardly also apply in the case of $\lambda \phi^4$ with $N=1$. This is because large $N$ is important when arguing that the dynamics is well-described by free fields with dynamically generated mass. The same logic applied when $N=1$ goes under the name of the Hartree approximation, and we see that if it were valid it would imply a late-time gaussian distribution along the lines argued above. 

But this differs sharply from standard arguments which, as described above, instead indicate a very nongaussian distribution of the form given in \pref{latenongauss}, and the direct comparison of how $n$-point functions evolve in \cite{TsamisWoodard} strongly suggest that the predictions of the Hartree approximation simply gives the wrong result. 

In order to handle this case we must broaden the scope of the formalism derived here to include the case when parameters like $m^2$ are themselves functions of background quantities, $m^2 = m^2(\bar \varphi)$, and so can evolve adiabatically over time if $\bar\varphi = \bar\varphi(t)$. In this case the vacuum probability, $\cP(\varphi)$, that is nominally time-independent can acquire a slow secular drift: $\cP = \cP(\bar\varphi,\varphi)$. 

When this is true the corresponding noise and drift parameters also inherit this background dependence, with for instance $\cF = \cF(\bar\varphi, \varphi)$ and so on. Should horizon exit be gaussian the dependence of $\cN$ and $\cF$ on $\bar\varphi$ is given by the above expressions with $q$ and $\nu$ regarded as functions of the instantaneous values of $m^2(\bar\varphi)$ {\em etc}, with gaussian exit in particular implying $\cN = \cN(\bar\varphi)$ is locally independent of $\varphi$ and $\cF = (q-\nu)(1-s-\epsilon) \varphi$ linear in $\varphi$. 

But $\varphi$ and $\bar\varphi$ are really just background and fluctuating components for a single field, $\Phi = \bar\varphi + \varphi$, and so functions like $\cF$ are really functions only of a single variable $\cF = \cF(\Phi)$ and not two separate quantities. The full function is determined from the above gaussian description by expanding about $\Phi = \bar \varphi$ and so $\cF(\bar\varphi + \varphi) \simeq \cF'(\bar\varphi) \varphi$. Using $m^2(\bar \varphi) = V''(\bar\varphi)$ we identify
\be
 \cF(\Phi) = \int^\Phi \exd \bar\varphi \, (q-\nu)(1-s-\epsilon) H\simeq \int^\Phi \exd \bar\varphi \left\{ \frac{V''(\bar\varphi)}{(3-\epsilon) H} + \cdots \right\} \,.
\ee
Notice this agrees with the usual result $\cF = V'/3H$ at leading order, as it must. 

There are also corrections to the standard expression involving powers of $\epsilon$ and $V''/H^2$. For example in the simplest case with $c_s$, $\epsilon$ and $H$ independent of $\bar \varphi$ and $V(\bar\varphi) = \frac{1}{4!} \, \lambda \bar\varphi^4$ the integral can be performed explicitly,
\bea \label{phi4drift}
 \cF (\Phi) &=& \left( \frac{3-\epsilon}{2} \right) H \int_0^\Phi \exd \bar \varphi \left[ 1 - \sqrt{1 - \frac{2 \lambda \bar \varphi^2}{(3-\epsilon)^2 H^2}} \right] \nn\\
 &=& \left( \frac{3-\epsilon}{4} \right) H\Phi \left\{ 2 -  \sqrt{1 - \frac{2 \lambda \Phi^2}{(3-\epsilon)^2 H^2}} - \frac{(3-\epsilon)H}{\sqrt{2\lambda} \; \Phi} \arcsin \left[ \frac{\sqrt{2\lambda} \; \Phi}{(3-\epsilon)H} \right] \right\}  \\
 &=&  \frac{\lambda \Phi^3}{6 (3-\epsilon)H} - \frac{ \lambda^2 \Phi^5}{20(3-\epsilon)^3 H^3} + \cdots \,,\nn
\eea
where we choose the integration constant so that $\cF(0) = 0$. When used in \pref{GenLateFP} --- together with the analogous expression for $\cN$ --- this formula modifies the late-time prediction for $\cP(\varphi)$ in a calculable (and IR safe) way. The subleading terms in \pref{phi4drift} also modify formulae such as \pref{phi2nstoch} at subleading order in $\lambda$, allowing them to be tested by precision higher-order calculations of $n$-point functions within the IR part of the field theory.\footnote{We do not see that the arguments of \cite{TsamisWoodard} exclude the existence of the higher-order corrections in powers of $\lambda$ we find above, and so it would be useful to sharpen the comparison to see if their existence can be tested using other tools.} 

\subsubsection*{Comparison with the dynamical RG}
\label{ssec:compDRG}

A closely related proposal for resumming late-time secular evolution \cite{OurDRG} (see also \cite{DRGcosmo}) is the dynamical renormalization group (DRG) \cite{DRG}. It is related because the essence of the resummation argument is in both cases a reliance on the existence of a broader domain of validity for the evolution equation of a quantity than on the perturbative steps that lead to its derivation. Consequently it is useful to compare how these procedures compare with one another in detail.

One case where such a comparison is possible is the large-$N$ limit of $\lambda \phi^4$ theory considered above. Ref.~\cite{OurDRG} shows that the resummation of the secular effects is in this case equivalent to a dynamical shift of the scalar mass, and indeed this comparison was made in order to test the DRG arguments against inferences drawn using the well controlled large-$N$ expansion. Given the above discussion of the large-$N$ case it follows that the DRG and stochastic methods also agree with one another when applied to this case. It is less clear whether there is agreement in the case $N=1$, largely because in this case it is not known how broadly the DRG resummation can be regarded as being equivalent to a dynamical mass shift.  

It would be instructive to have more comparisons of this type. However it was partly a dissatisfaction with our understanding of the systematics of the corrections to the DRG that led us to continue the search for a better framework, ultimately leading us to the formalism of Open EFTs \cite{OpenEFT} used here. 

\section{Discussion}
\label{section:discussion}

\subsection*{Summary of results}

In this paper, building on the results of \cite{OpenEFT}, we apply the tools of   open effective field theory to inflationary cosmology, with the aim to address  issues related with infrared singularities and resummation of  secular effects in inflation. Open EFTs allow us to find a master equation for a coarse-grained quantity built in terms of long-wavelength modes, that perturbatively accounts for information exchange  among long and short modes during inflation. 
 
To leading approximation the master equation for super-Hubble modes reduces to Starobinsky's formulation of stochastic inflation, in the form of a Fokker-Planck equation characterized by noise and drift functions, although we also find corrections to how these functions depend on system parameters like particle masses or slow-roll parameters. The master equation also has subleading contributions that go beyond Stochastic Inflation, such as those that decohere super-Hubble degrees of freedom due to their interactions with short wavelength  modes. 

All evidence so far supports the point of view that these master equation techniques lead to a consistent resummation of secular effects, and in this paper we test this by applying these tools to a three-parameter deformation of a massless spectator scalar in de Sitter space. In particular we compute the evolution and fluctuations of a spectator scalar of mass $m$ with time dependent sound speed $c_s$, within a power-law inflationary set-up (with constant slow-roll parameter $\epsilon$). This system is simple enough to be exactly solvable, but at the same time sufficiently rich to exhibit subtle IR singularities and secular effects when regarded as a perturbation to a massless field in de Sitter space. 

We obtain explicit expressions for the noise and drift functions characterizing the corresponding Fokker-Planck equation, and compute the corrections to its noise and drift as functions of the model parameters $m/H$, $\epsilon$ and $s = \exd \ln c_s \exd \ln a$. We find these corrections are just what is required to give IR safe expressions for the noise and drift, and so also to the late-time probability distribution $\cP(\varphi)$. This is a first step towards showing the IR safety of a wide variety of late-time observables. Scalar correlation functions are {\em not} similarly IR safe, but these IR singularities are driven by singularities in the {\em fluctuations} of the drift, rather than in singularities of the noise and drift functions themselves. 

We also generalize the Fokker-Planck equation to the case where the scalar mass is only locally gaussian at horizon crossing, with field dependent mass, and by doing so we obtain the late-time limit of massless spectator fields that self-interact through a $\lambda \phi^4$ interaction. The leading results agree with standard stochastic predictions, but seem to differ systematically at higher orders in $\lambda$.  

\subsection*{Future directions}

Open EFTs are likely to be useful to understanding the late-time limit for a number of different kinds of gravitational problems. Among those currently under study are the following.
\begin{itemize}
\item First, one can generalize our results for the corrections to the noise and drift to a broader class of models for which in the mass is field dependent. Besides accessing the $\lambda \phi^4$ case one might gain insight as to the late-time behaviour of fluctuations in moduli or the Higgs field in the very early universe. 
\item Making contact with observables requires moving beyond the spectator approximation to compute the scalar fluctuation variable $\zeta$, or, in a general gauge, the Sasaki-Mukhanov variable \cite{Mukhanov}, as well as of any isocurvature fluctuations. 
\item The generality of stochastic corrections and their ability to resum late-time behaviour can be tested by applying it to scenarios where explicit calculations are available. These include situations where other light fields are present during inflation, as electromagnetic spin one fields,  or fermions. There has been some study of stochastic versions of scalar QED, see \cite{Prokopec:2007ak}, Einstein-Maxwell systems \cite{Wang:2014tza} and the dynamics of minimally coupled fermions interacting with a scalar in de Sitter space \cite{Prokopec:2003qd} to which we hope our ability to systematize the stochastic framework can bring further insights, and against which predictions can be concretely tested.
\item It would be useful to go beyond the IR-finiteness of the late-time distribution function, $\cP(\varphi)$, to see if it can lead to something like a Bloch-Nordsieck theorem that can identify systematically IR safe quantities, and hopefully thereby to identify more systematically the theoretical errors in cosmological predictions. A bonus would be to be able efficiently to identify any large (but finite) `large logs' that capture the residual dependence of cosmological observable on large ratios of scale. 
\item Finally, as mentioned also in \cite{OpenEFT}, one might explore whether the Open EFT formalism has something useful to say for the information-loss problem in black hole physics. The issues arising there are similar to those in cosmology in that one follows only a subset of degrees of freedom, but there is information exchange between those that are tracked and those that are not. Furthermore, the oddities that are encountered occur at late times, and one seeks hidden reasons why EFT methods might fail in the late-time regime. (See \cite{BHsec} for a discussion of secular effects for black holes.)
\end{itemize}
All of these issues involve the late-time behaviour of open gravitating systems, and so are likely to profit from new insights obtained by bringing to gravity tools developed elsewhere for dealing with late-time issues.

\section*{Acknowledgements}

We thank Louis Leblond, Eugene Lim, Jamie McDonald,   Sarah Shandera, Martin Sloth, Alexei Starobinsky, Andrew Tolley, Nikolaos Tsamis,  Vincent Vennin, Matt Williams, Richard Woodard, and Mark Wyman  for useful discussions about stochastic inflation and secular and IR issues in inflation. The CERN TH Division, the Aspen Center for Physics, the Kavli Institute for Theoretical Physics (KITP) and the NYU Center for Cosmology and Particle Physics (CCPP) kindly supported and hosted various combinations of us while part of this work was done. This research was supported in part by the US DOE through Grant  No. DE-FG03-91-ER40682 as well as funds from the Natural Sciences and Engineering Research Council (NSERC) of Canada. Research at the Perimeter Institute is supported in part by the Government of Canada through Industry Canada, and by the Province of Ontario through the Ministry of Research and Information (MRI). Work at KITP was supported in part by the National Science Foundation under Grant No. NSF PHY11-25915. Work at Aspen was supported in part by National Science Foundation Grant No. PHYS-1066293 and the hospitality of the Aspen Center for Physics.

\appendix

\section{Fokker-Planck vs Schr\"odinger}

In general it need {\em not} be true that the Schr\"odinger equation,
\be \label{schro}
 i \dot \Psi =  \Bigl( - \kappa \, \nabla^2  + V  \Bigr) \Psi \,,
\ee
is equivalent to a Fokker-Planck equation,
\be \label{FPdefapp}
 \dot P = \nabla \cdot \Bigl( \cN \, \nabla P - \vec \cF \, P \bigr) \,,
\ee
with $P = |\Psi|^2$. Indeed inserting $\Psi = \sqrt{P} \; e^{iS}$ into \pref{schro} gives the pair of equations
\be \label{Rschro}
  \dot S = \frac{\kappa}{2} \, \nabla^2 \ln P  +\frac{\kappa}{4} (\nabla \ln P)^2 - \kappa (\nabla S)^2 - V \,,
\ee
and
\be \label{Ischro}
 \dot P = -2 \kappa \nabla \cdot \Bigl( P\, \nabla S \Bigr) \,,
\ee
rather than \pref{FPdefapp}. Both \pref{FPdefapp} and \pref{Ischro} tell us that $\dot P = - \nabla \cdot \vec J$ for a probability current $\vec J$, as required for $P$ to have a  time-independent normalization. Eq.~\pref{FPdefapp} therefore reproduces \pref{Ischro} whenever $\cN$ and $\vec\cF$ can be chosen so that \pref{Rschro} is consistent with the identification
\be
 -2\kappa \nabla S = \cN \nabla \ln P - \vec \cF \,.
\ee

A precise answer can be given for this in the case of a harmonic oscillator ({\em ie} $V = \frac12 \, k \,\vec x^2$) in $n$ dimensions (where $\vec x$ is an $n$-dimensional vector), for the class of states that are gaussian in $\vec x$.  (Notice this includes, but is not restricted to, the ground state, for which $S = -E_0 \,t$ and $\ln P = - \frac12 \, \alpha_0\,  \vec x^2$), for which \pref{Ischro} trivially tells us $\dot P = 0$ and \pref{Rschro} implies $E_0 =  \frac12 \, \kappa n \alpha_0 $ and $\kappa \alpha_0^2 = 2k$. But $\ln P = - \frac12 \, \alpha_0 \, \vec x^2$ and $\dot P = 0$ also solve \pref{FPdefapp} with $\vec \cF = - \nabla V = -k \vec x$ provided $\cN = k/\alpha_0$.)

In this gaussian case the Fokker-Planck equation captures the Schr\"odinger evolution for WKB states for which $S$ varies much more quickly than does $P$, since these states are `classical' to the extent that $-i \nabla \Psi \simeq \nabla S(x) \, \Psi$ shows that $\Psi(x)$ can effectively be regarded as both a position and momentum eigenstate. Whenever this is true there exists a gaussian classical distribution on phase space, $W(x, p)$, that reproduces the same mean and variance for both position and momentum that is predicted by $\Psi(x)$. (This is only possible because $\langle x_i p_j \rangle \simeq \langle p_j x_i \rangle$ in the WKB limit.) This then guarantees that the gaussian reduced distribution, $P(x) = \int \exd^n p \, W$, reproduces the variance and mean (for all $t$) for $\vec x$ predicted by $\Psi$, while the WKB relation $\vec p = \nabla S(x)$ ensures that this also dictates the evolution of the mean and variance of the momentum consistent with the evolution implied by $\Psi$ in the WKB limit. Since the coefficients $\cN$ and $\cF$ of the Fokker-Planck equation for $P$ are dictated by the evolution of $\langle x_i \rangle$ and $\langle x_i x_j \rangle$, we are then guaranteed they exist. 

For example, consider the simplest case 
\be
 \Psi = C \, \exp\left[ - \frac12 \, (A + i B) x^2 \right] \,,
\ee
with normalization constant satisfying $C^2 = \sqrt{A/\pi}$. The time-dependence of $A$ and $B$ is given by the Schr\"odinger equation, and determines the evolution of the mean and variance of the position and momenta from the formulae 
\be
 \langle x^2 \rangle = \frac{1}{2A} \,, \quad
 \langle p^2 \rangle = \frac{A^2 + B^2}{2A} \,, \quad
 \langle xp \rangle = \frac12 \left( i - \frac{B}{A} \right) \;\; \hbox{and} \;\;
 \langle px \rangle = \frac12 \left( -i - \frac{B}{A} \right) \,,
\ee 
so $\langle xp - px \rangle = i$ and $\langle xp + px \rangle = -B/A$. For this state the WKB limit (in which $S$ varies much faster than does $\ln P$) is given by $|B/A| \gg 1$, and this is a `squeezed' state \cite{squeezed} inasmuch as $\langle p^2 \rangle$ is much larger than $\langle x^2 \rangle$ in this limit. 

The classical phase-space distribution that captures this state is 
\be
 W = \tilde C \, \left[ - a x^2 - b p^2 - 2 c \, xp \right] \,,
\ee
for which normalization implies $\tilde C = \sqrt{ab-c^2}/\pi$. This predicts the variances
\be
 \langle x^2 \rangle = \frac{b}{2(ab-c^2)} \,, \quad
 \langle p^2 \rangle = \frac{a}{2(ab-c^2)} \,, \quad \hbox{and} \quad
 \langle xp \rangle =  \langle px \rangle = - \frac{c}{2(ab-c^2)}  \,,
\ee 
and so requiring these correctly reproduce $\langle x^2 \rangle$, $\langle p^2 \rangle$ and $\langle xp + px \rangle$ shows that $a$, $b$ and $c$ are given by
\be
 a = \frac{A^2+B^2}{A} \,, \quad 
 b = \frac{1}{A} \quad \hbox{and} \quad
 c = \frac{B}{A} \,.
\ee

The classical gaussian distribution for $x$ then is $P(x) = \int \exd p \, W(x,p)$, or
\be
 P(x) =  \sqrt{\frac{ab-c^2}{2\pi b}} \exp \left[ -  \left( a - \frac{c^2}{b} \right) x^2 \right] 
 = \sqrt{\frac{A}{\pi}} \; \exp \left[ -  A x^2 \right] \,,
\ee
in agreement with $|\Psi|^2$, and so $\nabla \ln P = P'/P = -\frac12 \, A x$ while $S = - \frac14 \, B x^2$ implies $\nabla S = S' = - \frac12 \, Bx$. Clearly the WKB classical regime corresponds to $|B/A| \gg 1$. 

In this example the Fokker-Planck equation captures the evolution implied by the Schr\"odinger equation provided only that it reproduces the right evolution for $A$, which requires $\cN = B/A$ up to terms subdominant in the WKB approximation. It is the noise that brings the news about the momentum variance, $\langle p^2 \rangle$,  to the Fokker-Planck equation (which nominally deals entirely with the classical statistics of $x$), because of the WKB relation $p = S' = - \frac12 \, B x$ which ensures $\langle p^2 \rangle \simeq B^2 \langle x^2 \rangle = B^2/2A$.

 \section{Calculation of fluctuations}
 \label{App:fluccalc}

This Appendix computes explicitly the fluctuations of a free massive spectator scalar field in power-law inflation, for use in deriving the corresponding Fokker-Planck equation and late-time evolution. We work in the Schr\"odinger picture starting from a wavefunctional, $\Psi[\varphi,\,t]$.  Our interest is in the time-evolution of the diagonal components of the density matrix, $\langle \varphi | \rho | \varphi \rangle = | \Psi[\varphi, t] |^2$ built from $\Psi$.

\subsection*{Action and hamiltonian}

Our starting point is the lagrangian density for a spectator scalar 
\be
  L = \int \exd^3 x\,a(t)^3\,\left[\frac12\,\dot{\phi}^2
  -\frac{c^2_s(t)}{2\,a^2(t)} \left(\nabla \phi \right)^2
  -\frac{m^2(t)}{2}\,\phi^2 \right] \,,
\ee
in an FRW spacetime with metric
\be
 \exd s^2\,=\,- \exd t^2 + a^2(t) \exd \vec{x}^2
\ee
and Hubble paramer $H(t)\,=\,\dot{a}/a$. Here $m(t)$ denotes the (possibly time-dependent) mass and $c_s(t)$ is a (possibly time-dependent) sound speed.

The Hamiltonian density in Schr\"odinger representation can be expressed in Fourier space as
\be
 \cH = \cH_0 + \sum_k \cH_k \,,
\ee
with $\cH_k$ for $k \ne 0$ given by 
\be
 {\cal H}_k\,=\,-\frac{1}{a^3}\,\frac{\delta^2}{\delta {\varphi}_k\, \delta {\varphi}_{-k}}+
  {a^3}\,\left[
 \frac{c_s^2\,k^2}{a^2}+ m^2
  \right] \,\varphi_k  \varphi_{-k}
\ee
where $\varphi^*_k \, = \,  \varphi_{-k}$. The contribution for the real zero-mode, $\varphi_0$, is
\be
 {\cal H}_0 \,=\,-\frac{1}{2\,a^3}\,\frac{\delta^2}{\delta {\varphi_0}^2}+\frac12 a^3\,
  {m^2}\,\varphi_0^2 \,.
\ee

\subsection*{Ground state wave functional}

We use this Hamiltonian to evolve the state wave-functional, $\Psi = \prod_k \Psi_k$, according to the Schr\"odinger equation,
\be \label{scheq}
   i\,\frac{\partial \Psi_k}{\partial t}\,=\,{\cal H}_k\,\Psi_k \,,
\ee
and for free fields we seek solutions subject to a gaussian ansatz,
\be
 \Psi[\varphi] = \prod_k \Psi_k[\varphi] \,=\, e^{-a^3(t) \, \beta_0(t) \,  {\varphi}_0}  \prod_{k}
 {\cal N}_k(t)\, \exp \Bigl\{ -a^3(t)\,\Bigl[ 
 \,\alpha_k(t)\,\varphi_k \,\varphi_{-k}
 \Bigr]  \Bigr\}
  \label{defPsi}
\ee
with ${\cal N}_k(t)$, $\alpha_k(t)$, $\beta_0(t)$ functions of $t$ now to be determined by substituting into \pref{scheq}. Notice the quantity $\beta_0$ here allows the possibility that the zero-mode has a nonzero mean in the ground state. 

We obtain in this way the following evolution equations for $\alpha_k$ and $\beta_0$:
\bea
  0&=& \dot{\alpha}_k+i \,\alpha_k^2+3\,H\,\alpha_k-i\left( \frac{c_s^2\,k^2}{a^2}+m^2\right)
 \hskip1cm {\rm for}\,\,k\ge0
\label{evea}
\\
  0&=& \dot{\beta}_0+ \left( 3 H+i \,\alpha_0 \right)\,\beta_0
\label{eveb}
\eea
where all quantities (including the Hubble parameter) can be time dependent, and the dot denotes derivative
with respect to time.  The additional equation for ${\cal N}_k$ ensures it evolves in a way that is consistent with normalization, but is not needed in what follows.

The eq for $\beta_0$ can be integrated to give
\be
   \beta_0(t)\,=\, \bar{\beta}_0 \left( \frac{a_0}{a(t)} \right)^3   \exp \left[ -i \,\int^t_0 \, \exd \tau \,\alpha_0(\tau)  \right]
\ee
where $\bar{\beta}_0 = \beta_0(t=0)$ is an integration constant fixed by initial conditions. It remains to find $\alpha_k$ by solving \pref{evea}. 

The solution for $\alpha_k$ can be made very explicit if we assume $c_s = c_0 (a/a_0)^s$, power-law expansion, $a = a_0 (t/t_0)^p$ (so that $H = p/t$ and $\epsilon = -\dot H/H^2 = 1/p$) and a time-independent ratio $m/H$. In this case equation \pref{evea} is integrated by changing variables to
\be\label{reak1}
  \alpha_k = -i \left( \frac{\dot u_k}{u_k} \right) = \-i\,a\,H\, \left[ \frac{\partial_a\, u_k(a)}{u_k(a)} \right] \,,
\ee
since \pref{evea} is then satisfied if $u_k$ solves the relevant Klein-Gordon equation,
\be \label{KGmodes}
  \ddot u_k + 3H \, \dot u_k + \left( \frac{c_s^2 k^2}{a^2} + m^2 \right) u_k = 0 \,.
\ee
For constant $\epsilon$, $s$ and $m^2/H^2$ this is solved by
\be
 u_k(a) = \tilde \cC_k \, y^q \,\sigma_k (y),
\ee
where $\tilde \cC_k$ is $a$-independent, provided $q$ and $y$ are chosen as
\be \label{qdef}
 q = \frac{3-\epsilon}{2\, (1-s-\epsilon)} \,,
\ee
and 
\be
   y(a,k) := \frac{1}{(1-s-\epsilon)}\left( \frac{c_s\,k}{a H} \right) = \frac{1}{(1-s-\epsilon)}\left( \frac{c_0\,k}{a_0 H_0} \right) \left( \frac{a_0}{a} 
   \right)^{1-s-\epsilon}   \,.
\ee 
The point of these changes of variables is that they turn eq.~\pref{KGmodes} into the Bessel equation for $\sigma_k$:
\be
  y^2\,\sigma_k''+y\,\sigma_k'+\left(y^2-\nu^2 \right)\,\sigma_k = 0 \,,
\ee
where primes here denote derivatives with respect to $y$. The order $\nu$ is given by
\be \label{eqfnu}
  \nu^2=\frac{1}{(1-s-\epsilon)^2}\left[  \frac{(3-\epsilon)^2}{4}-\frac{m^2}{H^2}\right] \,.
\ee

The solutions for $\sigma_k$ are (naturally) Bessel functions, and demanding agreement with the adiabatic vacuum before horizon exit tells us 
\be\label{bdsol}
  u_k \propto  \exp \left[ \mp i \int \exd t \left( \frac{c_s\,k}{a} \right) \right] \propto  e^{ \pm iy } \qquad \hbox{for $k/a \gg H$} \,,
\ee
of which we choose the lower sign since this turns out below to ensure the real part of $\alpha_k$ is positive (as required to ensure $\Psi_k$ can be normalized). This fixes the mode functions to be
\be \label{mods1}
  u_k (a) = \tilde \cC_k \, y^q(a,k) \,H^{(2)}_{\nu}\left[y(a,k)\right] =  \frac{\cC_k}{\sqrt{a^3 H}}  \; H^{(2)}_{\nu}\left[y(a,k)\right]
\ee
where $\cC_k \propto k^q \tilde \cC_k$ relabels the integration constants and $H^{(2)}_\nu$ the Hankel function of the second kind. The second equality in \pref{mods1} follows from eq.~\pref{qdef}, which implies $a^3Hy^{2q}$ is time-independent. Notice this reduces to the solution for a massive spectator field in de Sitter space in the limit $\epsilon \to 0$ and $s \to 0$. 

Although $\cC_k$ drops out of \pref{reak1} and (so does not contribute directly to $\alpha_k$), some later formulae are simpler if we choose $\cC_k$ so that the Wronskian, 
\be \label{Wronskiandef}
  \cW(u,v) := a^3 (u^* \dot v - v^* \dot u) \,,
\ee
satisfies $\cW(u,u) = i$. Among the formulae that simplify in this case is the expression for the real part of $\alpha_k$, as may be seen from
\bea
  \alpha_k+\alpha_k^*&=& -i \left( \frac{u_k^* \dot u_k - u_k \dot u_k^*}{|u_k|^2} \right) = \frac{1}{a^3\,|u_k|^2} \label{omefir}
  \\
   \hbox{and} \quad \alpha_k-\alpha_k^*&=&-i\,a\,H\,\left[\frac{\partial_a\,\left(|u_k|^2\right)}{|u_k|^2} \right] \,. \label{omesec}
\eea
Because $\cW$ is independent of time (when evaluated with solutions to \pref{KGmodes}) it is convenient to compute the implications for $\cC_k$ in the remote past, where $c_s k \gg aH$, in which case the Hankel function limit
\be
  H^{(2)}_\nu(y) \to \sqrt{\frac{2}{\pi y}} \;  e^{-iy +\frac{i\pi}{2} \left(\nu+\frac12 \right)} \quad \hbox{for $y \to \infty$} \,. 
\ee
can be used to infer
\be
 \left| \cC_k \right|^2 = \frac{\pi}{4(1-s-\epsilon)} \,,
\ee
for all $k$ and $\nu$. 

Consequently the quantity relevant to fluctuations in the main text is
\be \label{smp20}
    | u_k|^2  =  \frac{\pi}{4(1-s-\epsilon) a^3 H } \, | H_\nu^{(2)}(y) |^2  \,,
\ee
which with the asymptotic expression 
\be
  H^{(2)}_{\nu} \left(y\right)  \to \frac{i \Gamma(\nu)}{\pi}\,\left( \frac{y}{2} \right)^{-\nu} \quad \hbox{for $y \to 0$}  \,,
\ee
gives the small-$k$ limit
\be \label{smp2}
    | u_k|^2  \to \frac{2^{2\nu-2} |\Gamma(\nu)|^2 (1-s-\epsilon)^{2\nu-1}}{\pi a^3 H } \, \left( \frac{aH}{c_s k}
   \right)^{2\nu}  \,.
\ee
Finally, the case $\nu = \frac32$ is particularly simple because
\be 
 H_{3/2}^{(2)} (y) = \sqrt{\frac{2}{\pi y^3}} \;  \left( y - i \right) e^{-iy+i\pi} \,,
\ee
and so
\be
  u_k =- (1 - s -\epsilon)  \, \frac{H}{\sqrt{2 (c_s k)^3}} \; (y - i) e^{-iy} \qquad \hbox{for $\nu = \frac32$}
\ee
up to an irrelevant phase.

A further useful formula for later purposes is
\be
  a \, \partial_a y = -(1-s-\epsilon) \, y = -(1-s-\epsilon) \; k \, \partial_k y \,,
\ee
and so because $k^q u_k$ depends on $k$ and $a$ only through the combination $y(a,k)$ it follows that
\be
  a \, \partial_a \Bigl( k^{2q} | u_k|^2 \Bigr) = -(1-s-\epsilon) (k \, \partial_k ) \Bigl( k^{2q} | u_k|^2 \Bigr)  \,.
\ee

\subsection*{Evolution of the mean and variance}

Because the system is gaussian the Fokker-Planck equation is dictated by the evolution of the mean and variance, which we now compute using the above formulae. For each mode separately this is straightforward to do, starting with the probability density $P_k \,=\, \Psi^*_k \,\Psi_k$, which evaluates to
\be \label{fpdf1}
  P_k = \frac{a^3\left( \alpha_k+\alpha_k^*\right) \,
   }{\pi}\,\exp{\left\{-\frac{a^3}{ \left( \alpha_k+\alpha_k^*\right)} 
 \Bigl[ \left( \alpha_k+\alpha_k^*\right)\,\varphi_k -\delta_{k0}\,\beta_k\Bigr] \left[\Bigl( \alpha_k+\alpha_k^*\right)  
  \varphi_{-k} -\delta_{k0} \beta_k\Bigr]  \right\}} \,.
\ee
Translation invariance ensures the mean is only nonzero for the zero mode, which takes the value:
\be
  \langle \phi_0 \rangle = \int \exd \varphi_0 \, \varphi_0 \, P_0(\varphi_0) = \frac{\beta_0+\beta_0^*}{2\left(\alpha_0+\alpha_0^* \right)}
\ee
The two point function for the $k \ne 0$ modes (and the variance for the zero-mode around its nontrivial mean) similarly is
\be
  \langle \phi_k  \phi_k^* \rangle =  \int \exd \varphi_k \exd \varphi_k^* \Bigl[ \varphi_k \varphi_k^* \, P_k (\varphi_k , \varphi_k^* ) \Bigr]
   = \frac{1}{a^3\left(\alpha_k+\alpha_k^*\right)}  = |u_k|^2 \,.
\ee

For the Fokker-Planck equation, however, our interest is in the evolution of the coarse-grained position-space field, rather than the variance in any one mode. Proceeding as in \cite{OpenEFT} we define the coarse-grained field by 
%
%
\be
 \phi_\cS(r) =  
 \int \frac{\exd^3 k}{(2\pi)^3}\,
  {\cal S}\left[y(k) \right] \,
 \phi_k  \, e^{i k r } \,,
\ee
where ${\cal S}$ is a window function that projects out sub-Hubble modes, with the defining properties
\be \label{Slims}
  {\cal S}\left(y\right)  \to  1 \quad \hbox{for $y\ll1$} \qquad \hbox{and} \qquad
  {\cal S}\left(y\right) \to  0 \quad \hbox{for $y\gg1$} \,.
\ee
We evaluate statistical properties of the field $\phi_\cS$ using the joint probability distribution function $P  = \prod_k P_k$. 

Because momentum conservation only allows nonzero mean for $k = 0$ and because $\cS \to 1$ as $k \to 0$ the mean of $\phi_\cS$ is the same as for the zero-mode,
\be
  \langle \phi_\cS \rangle = \langle \phi \rangle = \frac{\beta_0+\beta_0^*}{2\left(\alpha_0+\alpha_0^* \right)} \,,
\ee
and so its time evolution is given by the equations of motion for $\alpha_0$ and $\beta_0$ as
\be
  \partial_t\,\langle \phi_\cS \rangle
  = \partial_t\,\langle \phi \rangle
  = i \left( \frac{\alpha_0 \beta_0^* - \beta_0 \alpha_0^*}{\alpha_0 + \alpha_0^*} \right) \,.
\ee
This is related in the expected way to the mean of the canonical momentum,
\be \label{piav}
 \langle \Pi_\cS \rangle = \langle \Pi \rangle = \int \exd \varphi_0 \; \Psi^* \left( -i \, \frac{\partial \Psi}{\partial \varphi_0}\right) = ia^3 \left( \frac{\alpha_0 \beta_0^* - \beta_0 \alpha_0^*}{\alpha_0 + \alpha_0^*} \right) \,,
\ee
so $a^3 \partial_t \, \langle \phi \rangle = \langle \Pi \rangle$.

The coarse-grained position-space two-point function is similarly
\be \label{varexp}
  \left \langle  (\phi_\cS - \langle \phi_\cS \rangle)^2 \right\rangle  =  \int \frac{\exd^3 k}{(2\pi)^3}\,|u_k\, {\cal S}|^2 \,,
\ee
and so its time dependence becomes
\bea  \label{eqf2}
  \partial_t \left \langle  (\phi_\cS - \langle \phi_\cS \rangle)^2 \right\rangle &=&  \int \frac{\exd^3 k}{(2\pi)^3}\,\partial_t\left(|u_k\,{\cal S}|^2\right)
   =  H \int \frac{\exd^3 k}{(2\pi)^3}\, a\, \partial_a\left(|u_k\,{\cal S}|^2\right)  \nn\\
  &=&  - \frac{H}{2 \pi^2} (1 - s - \epsilon) \int_0^\infty \exd k \, k^{3-2q} \, \partial_k \left(k^{2q} |u_k\,{\cal S}|^2\right)  \\
  &=&   \frac{H}{2 \pi^2} (1 - s - \epsilon) \left[ \lim_{k \to 0} \left( k^3 |u_k|^2 \right) + (3-2q) \int_0^\infty \exd k \, k^2 \, |u_k\,{\cal S}|^2 \right] \ \nn \\
   &=&   (1 - s - \epsilon) \frac{H}{2 \pi^2} \lim_{k \to 0} \left( k^3 |u_k|^2 \right) - (3s+2\epsilon)H\left \langle  (\phi_\cS - \langle 
   \phi_\cS   \rangle)^2 \right\rangle  \,.\nn
\eea
This uses the property that $k^{2q} |u_k\,{\cal S}|^2 $ depends on the variables $k$ and $a$ only through $y$ and so satisfies $a \, \partial_a = -(1-s-\epsilon) \, k \, \partial_k$, as well as eqs \pref{Slims} and \pref{varexp}. Notice in particular that \pref{eqf2} shows how the variance does not depend on the detailed shape of $\cS$, and only on its limiting forms.

Eq.~\pref{eqf2} can be regarded as a differential equation from which the time-dependence of the variance can also be extracted directly without evaluating mode sums explicitly. The equation to be solved has the form
\be
 \partial_t Y + \alpha(t) Y = X(t) \,,
\ee
with $Y(t) := \langle \left( \phi_\cS - \langle \phi_\cS \rangle \right)^2 \rangle$ and the identifications
\be
 \alpha(t) :=  (2q-3)(1-s-\epsilon) H(t) = (3s + 2\epsilon) H_0 \left( \frac{a}{a_0} \right)^{- \epsilon} 
\ee
and 
\be
 X(t) := (1-s-\epsilon) \, \frac{H}{2\pi^2} \, \lim_{k \to 0} \Bigl( k^3 |u_k|^2 \Bigr) 
 = X_0 \left( \frac{a}{a_0} \right)^{-3 + 2\nu(1 - s - \epsilon)}  \,,
\ee
where 
\be
 X_0 := \frac{ |2^\nu \Gamma(\nu) (1 - s - \epsilon)^\nu |^2}{(2 \pi)^3} \left( \frac{H_0}{c_0} \right)^{2\nu} \lim_{\mu \to 0} \left( \frac{\mu}{a_0} \right)^{3-2\nu} \,.
\ee

Integration gives the general solution
\bea
 Y(t) &=& \left\{ Y_0 + \int_{t_0}^t \exd \tau X(\tau) \exp\left[ \int_{t_0}^\tau \exd u \, \alpha(u) \right] \right\} \exp\left[ - \int_{t_0}^t \exd v \, \alpha(v) \right] \nn\\
 &=& \left\{ Y_0 + \int_{a_0}^a \exd u \left( \frac{X}{uH} \right) \exp\left[ \int_{a_0}^u \exd \hat u \left( \frac{\alpha}{\hat u H} \right) \right] \right\} \exp\left[ - \int_{a_0}^a \exd \tilde u \left( \frac{\alpha}{\tilde u H} \right) \right]  \,,
\eea
where $Y_0 = Y(t_0)$ is the variance at $t = t_0$. Inserting the known time-dependence of $\alpha$ and $X$ and performing the integrals then gives the general solution
\bea \label{Fasoln}
 Y(a) &=& \left\{ Y_0 - \frac{X_0/H_0}{(3-2\nu)(1-s-\epsilon)} \left[ \left( \frac{a_0}{a} \right)^{(3-2\nu)(1-s-\epsilon)} -1 \right] \right\}\left( \frac{a_0}{a} \right)^{(2q-3)(1-s-\epsilon)} \\
 &=& \left[ Y_0 + \frac{X_0/H_0}{(3-2\nu)(1-s-\epsilon)} \right] \left( \frac{a_0}{a} \right)^{(2q-3)(1-s-\epsilon)} - \frac{X_0/H_0}{(3-2\nu)(1-s-\epsilon)} \left( \frac{a_0}{a} \right)^{2(q-\nu)(1-s-\epsilon)} \,, \nn
\eea
where the two powers appearing in the last form are 
\be
 (2q-3)(1-s-\epsilon) = 2 \epsilon + 3s \,,
\ee
and
\be
 2(q-\nu)(1-s-\epsilon) = (3-\epsilon) \left[ 1 - \sqrt{1 - \frac{4m^2}{(3-\epsilon)^2 H^2}} \,\right] \approx \frac{2m^2}{(3-\epsilon)H^2} + \cdots \,.
\ee
It is this expression whose properties are explored in the main text.

\section{Other dispersion relations}
\label{app:ghostinf}

The formalism developed in the main text can be applied to models with non-standard kinetic terms as well. In this appendix we discuss the case of {\it ghost inflation} \cite{ArkaniHamed:2003uz}, where the Lagrangian density takes the form:

\be
L\,=\,\int d^3 x\,a(t)^3\,\left[\frac12\,\dot{\chi}^2 
-\frac{1}{2\,M^2\,a^4(t)} \left(\nabla^2 \chi \right)^2
-\frac{m^2(t)}{2}\,\chi^2
\right].
\ee

Here $M$ is a mass scale which we take to be constant here, though it could also be taken to be time dependent with $M(t)\slash H(t)$ constant as done in the main text and we continue to assume for the mass term $m^2(t)$ above. Also, just as for the cases treated in the main text, the dynamics of $\chi$ does {\em not} backreact on the geometry.  The fluctuations in this theory have a dispersion relation $\omega^2\propto k^4/M^2$, which could arise physically from situations where the scalar has vanishing sound speed, since then the terms with two spatial derivatives in the equations of motion would vanish. 

We perform the standard spatial Fourier mode decomposition of $\chi(\vec{x},t)$ and construct the Hamiltonian ${\cal H}_k$ for each mode

\be
{\cal H}_k\,=\,-\frac{1}{a^3}\,\frac{\delta^2}{\delta \hat{\chi}_k\, \delta \hat{\chi}_{-k}}+
{a^3}\,\left[ 
\frac{k^4}{M^2\,a^4}+ m^2
\right]\,\hat \chi_k \hat \chi_{-k}
\ee

An analysis following what was done in Appendix \pref{App:fluccalc} shows that the ground state wave-functional takes the form in eq.\pref{defPsi}, where 
the Schr\"odinger equation for each $\Psi_k$ now implies

\bea
0&=& \dot{\alpha}_k+i \,\alpha_k^2+3\,H\,\alpha_k-i\left( \frac{k^4}{M^2\,a^4}+m^2\right)
\hskip1cm {\rm for}\,\,k\ge0
\label{eveaghost}
\\
0&=& \delta_{k0}\left[ \dot{\beta}_0+ \left( 3 H+i \,\alpha_k \right)\,\beta_0\right]
\label{evebghost} 
\eea 

The relevant transformation for the Ricatti equation satisfies by $\alpha_k$ follows eq.(\pref{reak1})

\be\label{reak1ghost}
\alpha_k\,=\,-i\,a\,H\,\frac{\partial_a\,u_k(a)}{u_k(a)}
\ee
with 
\be
u_k \,=\,\frac{\cD_k}{\sqrt{a^3\,H}}\,\sigma_k\left(y\right)\,
\ee
and where we introduced the variable
\be
y\,=\,\frac{1}{(2-\epsilon)}\,\frac{k^2}{a^2 H^2}\,\frac{H}{M}
\ee 
In terms of these new variables, eq \pref{eveaghost} can be re-expressed as
\be
y^2\,\sigma_k''+y\,\sigma_k'+\left(y^2-\nu^2 \right)\,\sigma_k\,=\,0,
\ee
with primes denoting derivatives along $y$ and $\nu$ given by

\be \label{eqfnughost}
 \nu^2=\frac{1}{(2-\epsilon)^2}\left[
 \frac{(3-\epsilon)^2}{4}-\frac{m^2}{H^2}\right].
\ee 
The requirement of matching with the correct vacuum at early times uniquely fixes the solution and the integration constants, yielding

\be \label{mods1}
u_k\,=\,\frac{i\,\sqrt{\pi}}{2}\,
\frac{1
}{\sqrt{\left(2-\epsilon\right) a^{3}\,H }}\,H^{(2)}_{\nu}\left(
\frac{1}{(2-\epsilon)}\,\frac{k^2}{a^2 H^2}\,\frac{H}{M}
\right)
\ee
with $H^{(2)}_\nu$ the Hankel function of first kind.

Notice that
\be
\frac{\partial y}{\partial a}\,=\,-(2-\epsilon)\,\frac{y}{a}
\ee
so that  
\be \label{derpkghost}
\frac{\partial |u_k|^2}{\partial a}\,=\,-\left[ 3-\epsilon+y\,\left(2-\epsilon\right)\,\frac{\partial_y | H_\nu^{(2)}(y)|^2}{| H_\nu^{(2)}(y)|^2} \right]\, \frac{|u_k|^2}{a}
\ee
and that the real and imaginary parts of $\alpha_k$ are given by
\bea
\alpha_k+\alpha_k^*&=&\frac{1}{a^3\,|u_k|^2} \label{omefirghost}
\\
\alpha_k-\alpha_k^*&=&-i\,a\,H\,\left[\frac{\partial_a\,\left(|u_k|^2\right)}{|u_k|^2} \right]  \label{omesecghost}
\eea
In the limit of $k/(a H) \ll1$, using the
fact that
\be \label{lslim}
H^{(2)}_{\nu}\left(y\right)\,\to\,i\,\frac{\Gamma(\nu)}{\pi}\,\left( \frac{y}{2}\right)^{-\nu} \hskip1cm {\rm for}
\hskip1cm y\ll1
\ee
we can write

\be \label{smp2ghost}
| u_k|^2\,=\,\frac{2^{2\nu} \,\Gamma^2(\nu)}{4\,\pi\,a^3\,H}\,\left(
\frac{1}{(2-\epsilon)}\,\frac{k^2}{a^2 H^2}\,\frac{H}{M}
\right)^{-2\nu}
\hskip1cm,\hskip1cm  k/(a H)\ll1
\ee

\subsection*{The noise}

The new dispersion relation does not affect $\alpha_0$ so that using the fact that the noise term can be written generally as
\be
{\cal N}=\frac{i}{4\,\pi^2}\int_0^\infty d k\,\frac{k^2}{a^3}
\left(
\frac{\alpha_k-\alpha_k^*-\alpha_0+\alpha_0^*}{\alpha_k+\alpha_k^*}
\right),
\ee
we have that 
\bea
{\cal N} &=&\frac{1}{4 \pi^2}
\int_0^\infty\,k^2\,d k\,
\left\{
a\,H\,\partial_a\left( |u_k|^2\,{\cal S}\right)+H\left[3-2\nu(2-\epsilon)-\epsilon\right]\,  |u_k|^2\,{\cal S}
\right\}
\eea
Writing the integral in terms of the variable $y$ and using 
\be
k^2\,d k\,=\,y^\frac12\,(a H)^3\,\left(\frac{M}{H}\right)^\frac32\,\frac{(2-\epsilon)^\frac32     }{2}\,d y,
\ee
we obtain
\bea
{\cal N}&=&-\frac{(2-\epsilon)^\frac32}{8 \pi^2}\,{H^4\,a^3}\,\left( \frac{M}{H}\right)^\frac32
\int_0^\infty\,\sqrt{y}\,d y\,\left[ 
\frac{\partial \ln |  H_\nu^{(2)}(y) |^2 }{\partial \ln y}
+2 \nu\right]\,|u_k|^2
\nonumber\\
&=&-
\frac{(2-\epsilon)^\frac12}{32 \pi}\,{H^3}\,\left( \frac{M}{H}\right)^\frac32
\int_0^\infty\,\sqrt{y}\,d y\,\left[ 
\frac{\partial \ln |  H_\nu^{(2)}(y) |^2 }{\partial \ln y}
+2 \nu\right]\, |  H_\nu^{(2)}(y) |^2\nonumber
\\
&=&\frac{(2-\epsilon)^\frac12}{32 \pi}\,{H^3}\,\left( \frac{M}{H}\right)^\frac32
\left[
\left(
y^\frac32\,|  H_\nu^{(2)}(y) |^2
\right)_{y\to0}
+\left(\frac32-2\nu\right)
\,\int_0^\infty y^\frac12\, |  H_\nu^{(2)}(y) |^2\,{\cal S}(y)\nonumber
\right]\\
\eea
Thanks to eq \pref{lslim}, the previous integral can be re-expressed as 
\bea
{\cal N}&=&
\frac{(2-\epsilon)^\frac12}{32 \pi}\,{H^3}\,\left( \frac{M}{H}\right)^\frac32
\left[
\frac{\Gamma^2(\nu)}{\pi^2}\,2^{2\nu}
\left( 
\mu^{\frac32-2\nu}
\right)
+\left(\frac32-2\nu\right)
\,\int_\mu^\infty y^\frac12\, |  H_\nu^{(2)}(y) |^2\,{\cal S}(y)\nonumber
\right]\\
\eea
The quantity inside the square parenthesis is IR safe, and  well defined in the limit $\mu\to0$, as long as $\epsilon<2$, as we tacitly assumed so far.  

\subsection*{A special case}

There's a special case where things become simpler. In the limit $\epsilon\to0$, $m\to 0$,  we have $\nu\to3/4$, and the expression for the noise simplifies becoming

\bea
{\cal N}&=&
\frac{H^3}{4 \pi^2} \,\left( \frac{M}{H}\right)^\frac32\,
\left[
\frac{1}{2\pi}
\,\Gamma^2\left(\frac34\right)
\right]
\eea
Notice the correction with respect to standard result, that is weighted by $(M/H)^\frac32$.  The noise amplitude is enhanced if this
quantity is large.

\section{Large-$N$ scalars}
\label{app:largeN} 

We here briefly review the important points about $\lambda \phi^4$ theory in the large-$N$ limit. The system of interest consists of $N$ real scalar fields represented by the column vector $\Phi$, with lagrangian density
\be \label{app:Nlag}
  - \cL = \frac12 \, \partial_\mu \Phi \cdot \partial^\mu \Phi + \frac{\lambda}{4!} \, (\Phi \cdot \Phi)^2 \,.
\ee
Following \cite{Coleman} it is useful to define $g := \lambda N$ and introduce an auxiliary field $\chi$, through
\bea \label{app:Nlag2}
  - \cL &=& \frac12 \, \partial_\mu \Phi \cdot \partial^\mu \Phi + \frac{g}{4!N} \, (\Phi \cdot \Phi)^2 - \frac{3N}{2g} \left[ \chi_0 + \chi - \frac{g}{6N} (\Phi \cdot \Phi) \right]^2 \nn\\
  &=&  \frac12 \, \partial_\mu \Phi \cdot \partial^\mu \Phi + \frac12  (\Phi \cdot \Phi) (\chi_0 + \chi) - \frac{3N}{g} (\chi_0 \chi )- \frac{3N}{2g} \, \left( \chi^2 + \chi_0^2 \right)\,,
\eea
where integrating out $\chi$ in the first line returns the action to \pref{app:Nlag}, while the second line makes the large-$N$ limit most transparent. In these expressions $\chi_0$ represents the expectation value of $\chi$ and is determined by requiring a vanishing $\chi$ tadpole, giving
\be \label{app:tadpole}
 \chi_0 = \frac{g}{6N} \left \langle \Phi \cdot \Phi \right\rangle = \frac{\lambda}{6} \left \langle \Phi \cdot \Phi \right\rangle \,.
\ee

The utility of \pref{app:Nlag2} is twofold. First, after using \pref{app:tadpole} this representation shows that all factors of $\lambda = g/N$ are associated with $\chi$ propagators (which are also local in position space). Second, it shows that the integral over $\Phi$ is gaussian, describing $N$ scalars with mass $m_\phi^2 = \chi_0$ but without self-interactions, coupled linearly to the field $\chi$. In particular, this implies the standard calculation can be done to evaluate $\langle \Phi \cdot \Phi \rangle = 3N H^4/(8 \pi^2 m_\phi^2)$ (up to $1/N$ corrections), and so implies $\chi_0$ must satisfy
\be
 \chi_0 = \frac{gH^4}{16\pi^2 \chi_0} = \frac{\lambda N H^4}{16\pi^2 \chi_0} 
 \qquad \hbox{and so} \qquad
 \chi_0 = m_\phi^2 = \frac{\sqrt{g} \;H^2}{4\pi} \,.
\ee

In the large-$N$ limit where $g = \lambda N$ is fixed, we see the leading approximation drops $\chi$-exchange and so leaves a single free scalar whose mass is $m_\phi^2/H^2 = \sqrt{g}/4\pi$. Notice these statements do not require $g$ to be particularly small, though our applications to inflation require $\sqrt{g}$ to be at most order $4\pi$.

\end{document}